\documentclass[12pt, titlepage]{article} 
\usepackage[final]{graphics}
\usepackage{graphicx}
\usepackage{epstopdf}
\usepackage{setspace}
\usepackage{fullpage}
\usepackage{amsmath}

\begin{document}
\thispagestyle{empty}
\pagenumbering{roman}


\mbox{}\vspace{.8in}\mbox{}\\
\centerline{ \large \bf {A 233 km Circumference Tunnel for $e^+$$e^-$, $p$$\bar {p}$, and $\mu^{+} \mu^{-}$ Colliders}} 
\vspace{40mm}\\
\centerline{A Thesis}
\centerline{presented in partial fulfillment of requirements}
\centerline{for the degree of Master of Science}
\centerline{in the Department of Physics}
\centerline{The University of Mississippi}
\vspace{40mm}\\
\centerline{by}
\centerline{George Thomas Lyons III}
\centerline{December 2011}

\clearpage

\thispagestyle{empty}
\pagenumbering{roman}
\setcounter{page}{0}
\mbox{}\vspace{205mm}\\
\centerline{\large Copyright George T. Lyons III 2011}
\centerline{\large ALL RIGHTS RESERVED}

\clearpage

\section*{}
\addcontentsline{toc}{section}{ABSTRACT}
\pagenumbering{roman}
\setcounter{page}{2}
\mbox{}\vspace{.15in}\\
\centerline{\bf{ABSTRACT}}
\mbox{}\\
\indent 
In 2001 a cost analysis survey was conducted to build a 233km circumference tunnel in northern Illinois in which to build a Very Large Hadron Collider.  Ten years later I have reexamined the proposal, taking into consideration the technological advancements in all the aspects of construction cost analysis.  I outline the implementations of $e^+ e^-$, $p{\bar{p}}$, and
$\mu^+\mu^-$ collider rings in the tunnel using 21${\rm{st}}$ century
technology. The $e^+e^-$ collider employs a Crab Waist Crossing, ultra
low emittance damped bunches, 12 GV of superconducting RF, and 0.026
Tesla low coercivity grain oriented silicon steel/concrete dipoles. The
$p{\bar{p}}$ collider uses the high intensity Fermilab $\bar{p}$ source,
exploits high cross sections for $p\bar{p}$ production of high mass
states, and uses 2 Tesla ultra low carbon steel/YBCO superconductor
magnets run with liquid neon. The $\mu^+\mu^-$ ring ramps the
$p{\bar{p}}$ magnets at 8 Hz every 0.4 seconds, uses 250 GV of SRF, and
mitigates neutrino radiation with a phase shifting roller coaster FODO
lattice.  Such colliders could provide the means necessary for Chicago, and more importantly America, to stay  relevant  and competitive in the international marketplace of particle physics.

\clearpage

\section*{}
\addcontentsline{toc}{section}{DEDICATION}
\mbox{}\vspace{.15in}\\
\centerline{\large \bf DEDICATION}
\mbox{}\\
To Admiral Ackbar, the Rebel Alliance, and the United States of America. 
\newline Also to Dr. Donald Summers, a great mentor and friend.
\newline Pro scientia et sapientia.

\clearpage

\section*{}
\addcontentsline{toc}{section}{ACKNOWLEDGEMENTS}
\mbox{}\vspace{.15in}\\
\centerline{\large \bf ACKNOWLEDGEMENTS}
\indent

I would like to thank my committee chair, Dr. Donald Summers and the rest of my committee, Dr. Lucien Cremaldi and Dr. Luca Bombelli.  Also, many thanks to J. Scott Berg, Alain Blondel, Alakabha Datta, Murugeswaran Duraisamy, Henry Frisch, Steve Geer, Eliana Gianfelice-Wendt, Keith Gollwitzer, Sten Hansen, Terry Hart, Drew Hazelton, Tianhuan Luo, Steve Mrenna, Robert Palmer, Lalith Perera, Paul Rubinov, Tanaji Sen, and Alvin Tollestrup for enlightening conversations.
\vspace{.1in}\\
\indent
Above all, I would like to thank my Mom, Dad, Ms.\,Ann, and Jessy, for continued support throughout all my endeavors in life, both academic and personal.  I couldn't have done it without you.

\clearpage

\vfill\eject
\mbox{}\vspace{.0in}\
\begin{singlespace}
\tableofcontents
\end{singlespace}
\pagebreak

\vfill\eject

\section*{}
\addcontentsline{toc}{section}{LIST OF TABLES}
\mbox{}\vspace{.0in}\
\begin{singlespace}
\listoftables
\end{singlespace}
\pagebreak

\vfill\eject

\section*{}
\addcontentsline{toc}{section}{LIST OF FIGURES}
\mbox{}\vspace{.0in}\
\begin{singlespace}
\listoffigures
\end{singlespace}
\pagebreak

\pagestyle{plain}
\pagenumbering{arabic}
\setcounter{page}{1}

\mbox{}\vspace{.0in}\\
\section*{Introduction}
\addcontentsline{toc}{section}{Introduction}
\indent

For many years the University of Mississippi has 
collaborated on experiments in the Fermilab fixed 
target\,\cite{Anjos,Aitala} and collider\,\cite{Abazov2009,Abazov2009B} 
programs. In this thesis, we will explore the design of a large, 233 km 
circumference multipurpose ring for energy frontier colliders, utilizing more cost effective machines and methods. The colliders use several new technologies developed and refined in the last decade. 

Advances in the automation of some, but not all, aspects of tunnel boring 
machines and wall reinforcement are explored.  Improvements to the tunnel's structural integrity will also be discussed.

For the $e^+ e^-$ collider very low emittance bunches are combined 
with a relatively large crossing angle and a short vertical focal 
length at the interaction point for high luminosity by utilizing new methods of focusing collision regions implementing the Crab Waist crossing design. Also, we can integrate the use of low coercivity, grain oriented silicon steel for dipoles.

A $p \bar{p}$ hadron collider is chosen because the cross section for 
many high mass objects is much larger than for $p p$ collisions. Single 
bore iron dipoles permit high temperature superconductor ribbons running 
at liquid neon temperatures. Advances have been made in $\bar{p}$ 
production at Fermilab. A detector with more physics and fewer 
background interactions per crossing requires fewer channels and will consequently cost 
less.

The same dipoles used for $p \bar{p}$ can also be ramped to accelerate muons
to almost 20 TeV with less RF than the proposed International Linear Collider.
The ramping rate is slow enough to permit superconductors.
A simple method which solves the neutrino radiation problem will be worked out.
A 40 TeV center of mass muon collider with an effective energy reach
30 times higher than the CERN LHC is envisioned to do
physics\,{\cite{Barger,Eichten} at Fermilab.

Chicago is one of the great cities in the world and it needs to be home to one of the greatest scientific instruments in the world to maintain its position on a competitive planet. 

\clearpage



\mbox{}\vspace{.0in}\\
\section{Boring a 233 km Long Tunnel: 
\newline A Construction Cost Analysis}

\subsection{Drilling}
\label{sec:meaningfulname}
\indent

Essentially in the world of major underground drilling there are three major techniques commonly employed.  These methods include tunneling machinery that implements directional drilling bits, microtunnel boring machines (MTBM), and tunnel boring machines (TBM).  The most basic and commonly used of the three is directional drilling.  Applications of directional drilling are vast, but seen mostly in oil well preparation.  The large advantages of directional drilling techniques include a relatively low cost of operation (as it is unmanned), a multi directional drilling path, and the most easily excavated milling bi-product (a fine silt/water mixture referred to as ``muck'').  Also the need to change out the tunneling bit head has greatly reduced as synthetic diamond tipped and diamond ceramic bit heads have become increasingly available.  The overwhelming disadvantage of directional drilling however, comes in the form of a very limited drill head diameter, topping out at approximately 900mm; however, this might still be large enough for access shafts. For the purposes of this paper, we will focus most directly on tunnel boring machines (TBM) and microtunnel boring machines (MTBM).  

Like directional drilling, microtunnel boring machines are most commonly operated through remote control (thus largely unmanned).  While this is also a great advantage to the cost of operation and efficiency of a tunneling operation, MTBMs also share the limitations of directional drilling in the fact that they too have dimensional boring size restraints. While MTBMs are definitely larger than directional drilling methods, topping out at approximately 1.5 meters (~5 ft), there would still not be enough space for manned transportation within the underground collider ring.  Also, most conventionally MTBMs incorporate a pipe jacking mechanism that essentially drives a preassembled pipe into the tunnel to reinforce the walls of the tunnel.  Problems occur in this model as after a few kilometers of pipe jacking, friction does not allow for further pipe lengths to be installed.  However, this tunneling method should not be completely discounted, as it might be useful for installing larger elevator shafts.

Tunnel boring machine methods are by far the most useful methods introduced throughout this section.  TBMs are the most costly method of underground tunneling excavation, but are also the most sizable.  Often referred to as a ``mole'', TBMs have large circular faces that rotate and dismantle any solid terra in its path. By utilizing a pressurized mechanism that breaks away the rock face while allowing the large chunks of rock to fall through openings in its rotating face, TBMs are extremely efficient in burrowing through and disposing of solid surfaces.  One obvious advantage of breaking apart the tunneling face into larger chunks of particulate is the vast amount of energy saved in not grinding and refining the particulate down into a finer milled product.  In acquiring larger chunks however comes the problem of disposing of a larger milling byproduct.  These larger pieces of rock typically have to be disposed of by an integrated conveyor belt system that moves the rock away as it falls through the grinding mechanism face.  Although there are many methods of reinforcing the tunnel while the drilling continues throughout its directed path, most methods incorporate some combination of drilling pilot holes in the walls of the tunnel and placing reinforcement bolts with wire mesh, solid walls, concrete matrices, etc. to eliminate the possibility of wall collapse. The biggest advantage of utilizing TBM methods comes in the sheer size of the tunnel TBMs can accomplish.  Though most TBM faces are 3-5 meters ($\sim$10-16 feet) in diameter, current TBM manufacturing companies are pushing nearly 16-meter ($\sim$52 feet) production models.  

\subsection{Past Cost Estimation}
\label{sec:meaningfulname}
\indent

In 2001 an initial cost estimation was projected for the construction of a very large hadron collider tunnel (VLHC) to be constructed in Northern Illinois for the Fermi National Accelerator Laboratory.  This report was prepared by CNA Consulting Engineers and prepared an array of building initiatives and a very comprehensive cost analysis for several scenarios in which the collider tract could be constructed\,\cite{CNA, Lach}.  The final cost projection put the entire project in the \$2.5-2.8 billion dollar range, depending on tract orientation and location.  All of the intrinsic costs will be further explored in the following paragraphs.

One of the most fundamental parts of estimating the cost of tunneling underground comes in a simple equation that determines the quality (Q) of the material being drilled.  Quality is most directly correlated to the hardness and softness of the rock being drilled.  The harder the rock, the higher the Q factor, thus the better the quality.  Conversely, as rock is softer, its Q goes down, and its relative quality is considered bad. Typically the range of Q is set somewhere between .2 and 50 and is determined by the following equation:

\begin{equation}
Q = {{RQD} \over {J_n}} \times {{ J_r} \over {J_a}} \times {{ J_w} \over  {SRF}}
\label{tray}
\end{equation}

Where from Eq.\ref{tray} , RQD = Rock Quality Designation ,  $J_n$ = Joint set number,  $J_r$ = Joint roughness number,  $J_a$ = Joint alteration number,  $J_w$ = Joint water reduction number, and  SRF = Stress Reduction Factor, respectively.  Although all these variables are representative of physical characteristics of the rock's internal physical composition, the overall Q factor really reflects the structural rigidity and integrity of the rock to resist collapsing on itself.  The most cost efficient option for building this collider tunnel came in the form of an inclined track built in the northern portion of the state.  The cost effectiveness of this specific projection came in the limited range of Q quality factor ($\le$ 33) and the effects of inclining the track itself, so that major bedrock drilling could be put to a minimum. The projected range of  Q quality for this tunnel to be pushed through the northern incline ranged from $\sim$.33-33.  Overall, this ranging factor essentially represents the number of reinforcement measures needed per unit of tunnel.  For example, in the low end of the spectrum (0.33), reinforcement rock bolts would need to be drilled every 1.5 m and the thickness of the shotcrete wall covering the tunnel surface would need to be around 0.22 m thick.  Conversely, on the other end of the Q factor spectrum, the most rigid well ($\sim$33) could be composed of rockbolt spacings approaching 2.6 m and shotcrete thicknesses of around 0.07 m.  So it is easy to see that the integrity of the wall itself plays a large role the construction cost of the entire project.

It is also worth noting the cost per square foot of cross section of each projected tunnel construction in past projections.  When considering the 12 ft diameter tunnel at a circumference of 233 km, roughly 2,446,500 $\rm{m}^3$ of tunneling space would be created.  If we extended the interior tunnel diameter to 16 ft, at the same circumference, roughly 4,350,000 $\rm{m}^3$ of tunneling space would be created.  So by extending the interior diameter of the tunnel from 12 ft to 16 ft you would essentially double the volume of the tunnel; however, the cost of the 16 ft tunnel is not double the 12 ft.  In fact the 12 ft construction cost came in at approximately \$2.42 billion, while the 16 ft came in at \$2.71 billion.  A difference of approximately 10\% cost difference compared to a 100\% bigger tunnel.   Roughly 30\% of the total construction costs come from contingency costs.  A large part of this is the uncertainty of the actual costs of manning the mole operation.

\begin{figure}[ht]
  \centerline{\includegraphics[width=4in]{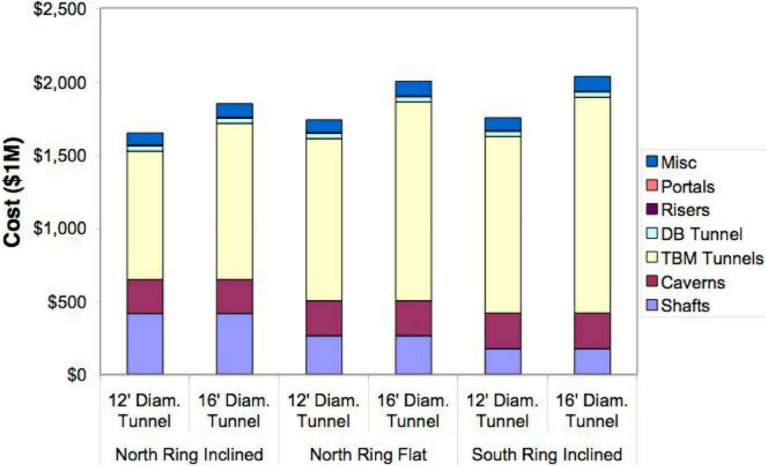}}
  \caption{Cost Analysis of Various Tunnel Diameters and Orientations\ \cite{CNA}}
  \label{cost}
\end{figure} 

\subsection{Future Projections}
\label{sec:meaningfulname}
\indent

Throughout examining the past projections of construction costs, many improvements in drilling technologies and cost effectives have arisen.  In the past decade great advancements in unmanned drilling possibilities have become evident, as well as the overall increases in diameters and efficiencies of directional drilling techniques, TBMs, and MTBMs.  Some of these advancements will be addressed in the following section as well as some new construction techniques that could serve for another building model altogether.

One such construction tactic could be the implementation of periodic reinforcement bulkheads throughout the shotcrete lined tunnel.  As it stands in past projections, shotcrete thicknesses ranged from 0.07$\rightarrow$0.22 m in the lining of the completed tunnel.  This thickness differential is essentially varying by a factor of 3 and if it could be reduced to a thickness of ~0.07 throughout the entire tunnel, roughly 200,000 $\rm{m}^3$ of shotcrete could be saved. This might be accomplished by installing period steel bulkheads to reinforce the structural integrity of the tunnel, much like the construction of modern day submarines.  The structural rigidity of a thin walled cylinder to resist external pressure increases by enclosing the ends. The resistance to collapse is derived from the equation for collapsing pressure,\ \cite{Marks},
  
\begin{equation}
W_c = K \times E \times {({{t} \over  {D}})^3}\ \rm{psi}
\label{tray1}
\end{equation}
 
Where from Eq.\ref{tray1}, K is a numerical coefficient that contains the stress variable, E is the elasticity modulus of the material, D = outside-shell diameter, and t = the thickness of the shell.  The modulus of elasticity of steel is $\sim$200 GPa compared to $\sim$20 GPa for concrete. Adding steel bulkheads periodically, as in a submarine, to the thin shotcrete lined cylindrical tunnel would save a lot of unnecessary shotcrete and add a lot more structural integrity to tunnel.  Also, the idea of creating these bulkheads using concrete might seem more feasible, but if you consider a 0.5 m thick steel bulkhead versus a 0.25 m thick concrete bulkhead, the steel bulkhead would be 80 times more resistant to buckling.  Obviously this would mean far fewer bulkheads, thus construction costs would be further reduced. 

The advancements in TBM automation are also very interesting when considering the construction cost of the VLHC in Northern Illinois.  Of the \$2.418 billion dollar construction budget for the 12' diameter northern inclined track, roughly 30\% came from a contingency budget.  A large percentage of this portion of the budget arises from the sheer cost of manned operation and liability of the TBM.  With great advancements in TBM automation, many TBMs are now operated by relatively few operators.  When considering the action of a TBM, four distinct operating actions must be considered: 1) The guidance of the TBM, 2) The setting of stabilizer supports on the tunnel walls for which the TBM establishes forward pressure, 3) The setting of wall reinforcement measures, and 4) The removal of rock and particulate and their disposal.  Through the great advancements in robotic controlling technology over the past decade, guidance mechanisms can be mostly operated from the surface, with the occasional subsurface manned intervention.  The setting of stabilizer legs for the forward pressure of the TBM are now completely automated by the TBM itself.  Essentially, they're just huge retractable legs at the rear of the TBM that continuously retract and reset as the TBM moves forward.  The setting of tunnel wall reinforcement has become almost completely mechanized and automated as well.  TBMs can now drill and place rock bolts periodically as needed and secure reinforcement mesh.  The only reinforcement tactic that still requires manned operation is the shotcrete application, and it is being rapidly addressed with new automated machinery that might function as a shotcrete robot, that would also be unmanned.  While rock and particulate disposal still need some manned assistance,  long conveyor tracts may be attached to the rear of the TBM to pull rock back to a more secure section within the tunnel.

Further investigation could be made into the construction cost analysis of the VLHC proposed almost a decade ago.  With the continued advancement in TBM automation, maybe soon we could see a machine that can bore 8 ft diameter tunnels (a very usable size) without much human interference or manned guidance at all. Other methods might be found to minimize costs on a 8 ft diameter tunnel's construction, as an 8 ft tunnel has roughly one-fourth the volume of a 16 ft tunnel.      

\clearpage

\mbox{}\vspace{.0in}\\
\section{Electron-Positron}

\subsection{Theory}
\label{sec:meaningfulname}
\indent

Electron-positron collision can result in the annihilation of each particle and the release of a pair of gamma rays, Eq.\ref{tray2}. 

\begin{equation} 
e^{+} + e^{-} \rightarrow \gamma + \gamma
\label{tray2}
\end{equation}

However if both particles have appreciable energies that are at or beyond the mass of the carriers of the weak force, W and Z bosons, then the weak force becomes comparable with that of electromagnetism and boson production is possible.  Although currently the heaviest particles realized in this annihilation are  $W^{+} - W^{-}$ pairs, the heaviest single particle is the Z boson.  This highly energetic annihilation is the driving force behind the proposal for the International Linear Collider, ILC, as it is likely the source of the much sought after Higgs boson, $(H)$, Fig.\ref{higgs}.

\begin{figure}[ht]
  \centerline{\includegraphics[width=6in]{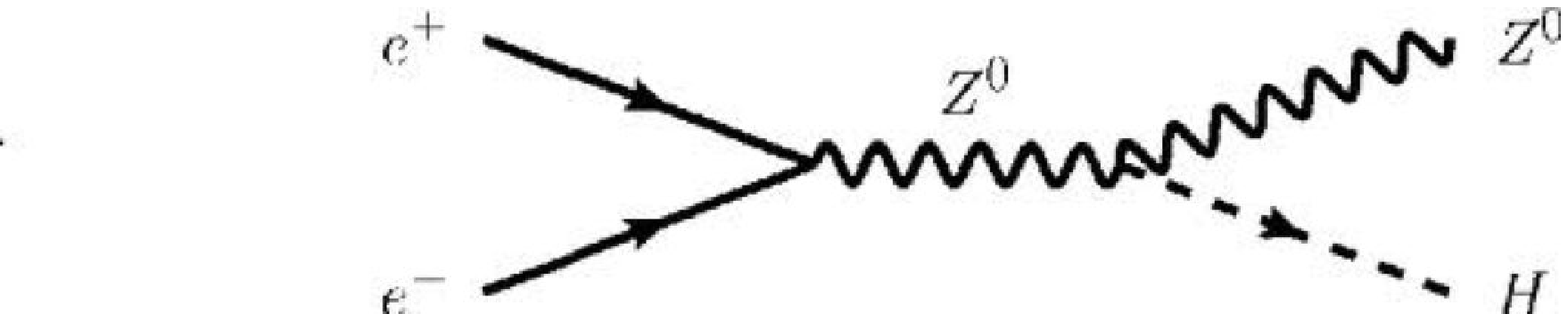}}
  \caption{Feynman Diagram of Higgs Boson Production.}
  \label{higgs}
\end{figure} 

\subsection{Past Research}
\label{sec:meaningfulname}
\indent

One of the most relevant research endeavors to date was the Large Electron-Positron Collider, LEP, conducted in 4 experiments at CERN from the early 90's to the early 2000's~\cite{Schopper}.   LEP allowed for the production of the Z boson, as well as the pair production of $W^{+} - W^{-}$ bosons.  In 2002, Tanaji Sen and Jim Norem introduced a proposal that would construct a Very Large Lepton Collider, VLLC, inside the proposed 233 km collider ring space built in conjunction with Fermilab \cite{Sen}.  The parameters surrounding the proposed collider are very interesting and the theoretical explanations for this construction will serve as the basis for this section.  

Examining the luminosity, $L$, of a specific Electron-Positron bunch collision, Eq.\ref{tray3} 

\begin{equation} 
L =  {1\over{2er_{e}}}{\xi_{y}\over\beta^*_{y}}\gamma I
\label{tray3}
\end{equation}

Where $\xi_{y}$ in the vertical beam-beam tune shift, $\beta^*_{y}$ is the focal length in the $y$ plane, $\gamma$ is the relativistic scaling factor (that essentially scales as the Collision Energy), and $I$ is the total beam current in both beams. Now if you consider the parameters examined at LEP in 1999 and the parameters proposed for the VLLC in Table \ref{table:nonlin}, the increase in luminosity is easily seen.  Essentially you get an increase in the vertical beam-beam tune shift,  $\xi_{y}$, of a factor of 2.28, a decrease in the focal length $\beta^*_{y}$ of 5, an increase in the relativistic $\gamma$ of 2.05, and an increase of the total beam current in a both beams, $I$, of about 3.95.  The adjusted parameters give you an overall increase in luminosity of 90 times the original luminosity of LEP, Eq.\ref{tray4}

\begin{equation} 
L_{VLLC} =  {1\over{2er_{e}}}{(2.28)\xi_{y}\over(.2)\beta^*_{y}}(2.05)\gamma(3.95) I = (92.31) L_{LEP} = 8.8\times 10^{33}\ \rm{cm^{-2}s^{-1}}
\label{tray4}
\end{equation}

Although this is a great increase in the overall luminosity of the Electron-Positron collisions compared at LEP, we can do better and these improvements will be the focus of our next section.

\subsection{Future Projections}
\label{sec:meaningfulname}
\indent

One of the most prominent and cutting edge techniques to improve electron-positron collisions comes in the form of creating a Crabbed Waist (CW) crossing at the intended collision site, Fig.\ref{crab}, \cite{Raimondi2007,Zobov,Shatilov}.  Instead of having the two beams collide head on, they are angled so that the collision only occurs in the section where they overlap.  Now the angling of beam collision alone would produce a great decrease in luminosity, as it would decrease the overlapping collision area.  Actually, it would go down exactly by a factor correlating to the angle at which the beam is tilted, or more commonly known as the Piwinkski angle, $\phi$, where $\phi$ is

\begin{equation} 
\phi =  {{\sigma_z}\over{\sigma_x}} \theta
\label{tray5}
\end{equation}

\begin{figure}[ht]
  \centerline{\includegraphics[height=3in, width=4.5in]{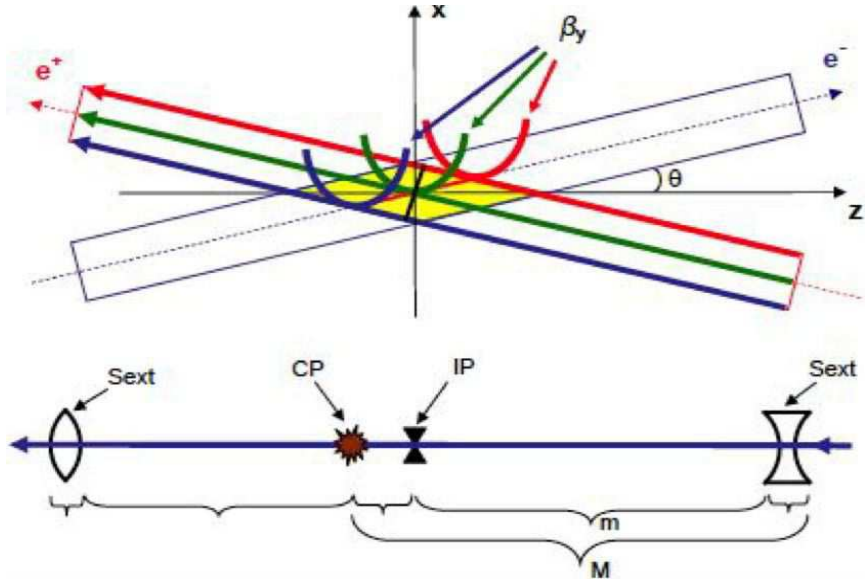}}
  \caption{Crabbed Waist Crossing Scheme\ \cite{Raimondi2007}}
  \label{crab}
 \end{figure}

However, a few more parameters could be tweaked to compensate for the luminosity loss and even improve it overall.  One of the most important consequences would be the ability to lower the vertical focal length at the collision point, $\beta^*_y$, to fit the decreased overlapping area.  Before we explore any changing of parameters in the luminosity equation though, let's explore a more direct approach approximating the luminosity parameters put forth by Sen and Norem \cite{Sen}.

Well taking the equation for luminosity to be Eq.\ref{tray6},

\begin{equation} 
L =  {{N_1N_2f_0}\over({4\pi \sigma_x \sigma_y})} 
\label{tray6}
\end{equation}  

The equation for tune shift, $\xi_y$, goes as Eq.\ref{tray7},

\begin{equation} 
\xi_y =  {{r_eN_2 \beta^*_y}\over({2\pi \gamma \sigma _y (\sigma_x + \sigma_y}))} 
\label{tray7}
\end{equation}

Then we can combine Eq.\ref{tray6} and Eq.\ref{tray7} to get Eq.\ref{tray8},

\begin{equation} 
L =  {\gamma \xi_y N_1f_0 (1 + {\sigma_y \over \sigma_x})\over({2r_e \beta^*_y)}} 
\label{tray8}
\end{equation}

The next step after resolving these equations is to simply put forth Eq.\ref{tray9}\ \cite{Raimondi2006},

\begin{equation} 
L =  {({2.167 \times 10^{34}) \times E(\rm{GeV}) \times {I}(A) \times \xi_y}\over({\beta^*_y(\rm{cm}))}} 
\label{tray9}
\end{equation}

But where did this constant come from? For the sake of sanity we'll calculate this constant directly. For a 1 GeV electron, ${\gamma} = {{10^9}\over511,000} = 1957$.  Taking the classical radius of an electron to be ${r_e} = 2.818 \times 10^{-13}\ \rm{cm}$ and the electron charge to be $1.602 \times 10^{-19}$ C, then we can calculate the constant to be Eq.\ref{tray10},

\begin{equation} 
{{{\gamma}\over (2 \times r_e \times q_e)} = {{1957}\over (2 \times 2.818\times 10^{-13} \times 1.602 \times 10^{-19})} = 2.167 \times 10^{34}} 
\label{tray10}
\end{equation}

plugging this into the parameters outlined in Sen and Norem, Table \ref{table:nonlin}, we get Eq. \ref{tray11},

\begin{equation} 
L =  {({2.167 \times 10^{34}) \times (200\ \rm{GeV}) \times (0.0114A) \times 0.18}\over({1\ \rm{cm})}} = 8.8 \times 10^{33}\ \rm{cm^{-2}s^{-1}}
\label{tray11}
\end{equation}

This result matches the result in Sen and Norem, Table \ref{table:nonlin}, almost exactly \cite{Sen}.  So now we can be confident to plug a few more recent parameter values established by more recent research endeavors to further our luminosity approximations.  One such parameter comes from International Linear Collider Reference Design Report, in which a focal length of $\beta^*_y$ =0.06 cm is achieved \cite{Phinney}.  Plugging this into our current parameters for Eq.\ref{tray9}, we get Eq.\ref{tray12} 

\begin{equation} 
L =  {({2.167 \times 10^{34}) \times (200\ \rm{GeV}) \times (0.0114A) \times 0.18}\over({0.06\ \rm{cm})}} = 1.5 \times 10^{35}\ \rm{cm^{-2}s^{-1}}
\label{tray12}
\end{equation}

This is astonishing, as luminosity has gone up by a factor of 17; however, to stay in line with current ILC specs, we should really explore the implications of going up to a 250 GeV collision energy. One consequence of going up to a higher energy is an increased allowable tune shift ($E: 200 \rightarrow 250 \ GeV$, corresponds to $\xi_y: 0.18 \rightarrow 0.23$).  However, we must consider the 100 Megawatt synchrotron radiation limit, so then the current, $I(A)$, scales accordingly, $({250\over200})^4$ = 2.44. The current then goes down by a factor of 2.44, ($I(A): 0.0114 \rightarrow 0.00467$).  Plugging in all these new parameters in Eq.\ref{tray9}, we get Eq.\ref{tray13}

\begin{equation} 
L =  {({2.167 \times 10^{34}) \times (250\ \rm{GeV}) \times (0.00467A) \times 0.23}\over({0.06\ \rm{cm})}} = 9.7 \times 10^{34}\ \rm{cm^{-2}s^{-1}}
\label{tray13}
\end{equation}

Which is still about 10 times higher than the luminosity from Eq.\ref{tray25} and 5 times higher than that of the ILC.  Another parameter that needs to be considered are the emittances that correspond to the crab waist scheme and their respective collision energies.  If we apply the crab waist scheme with half crossing angle $\theta$ = 17 mrad, assuming a bunch length $\sigma_z$ = 6.67 mm, and an overlap length $l$ = .5 mm = ${2\sigma_x}\over \theta$, then $\sigma_x$ = 4.25 $\mu$m.  If we know the focal length in the horizontal plane to be $\beta^*_x$ = 20 000 $\mu$m, then the emittance in the horizontal plane goes as 

\begin{equation}
{\epsilon_x} = {{\sigma_x^2}\over{\beta^*_x}} = 0.9\ \rm{nm}.
\label{tray14}
\end{equation}

Then calculating the $\delta{N_2}$ term,

\begin{eqnarray}
{\delta{N_2}} & = & {{2N_2 \sigma_x}\over{(\theta \sigma_z)}}\nonumber\\
{\delta{N_2}} & = & {{2\times (4.85\times 10^{11}) \times 4.25\ \mu{\rm{m}}}\over{(.0017 \times 6.67\ \rm{mm})}}\nonumber\\ 
{\delta{N_2}} & = & 3.63 \times 10^{10}
\label{tray15}
\end{eqnarray}

Now we can calculate the collision frequency, $f_0$, for 250 GeV,

\begin{eqnarray}
{f_0} & = & {{(Number\ of\ Bunches/Beam)\times c}\over{(Circumference\ of\ Collider)}}\nonumber\\
{f_0} & = & {{(46) \times (3\times10^5\ {\rm{km\over s}})}\over{(233\ \rm{km})}}\nonumber\\ 
{f_0} & = & 5.92 \times 10^{4}\ \rm{Hz}
\label{tray16}
\end{eqnarray}

Now we can calculate the beam size in the vertical plane, $\sigma_y$, for 250 GeV,

\begin{eqnarray}
{\sigma_y} & = & {N_1\delta{N_2}f_0}\over{4\pi \sigma_x L}\nonumber\\
{\sigma_y} & = & {(4.85\times 10^{11})(3.63 \times 10^{10})(5.92 \times 10^{4}\ \rm{Hz})}\over{4\pi(4.25\times 10^{-4}\ \rm{cm})(1.5 \times 10^{35}cm^{-2}s^{-1})}\nonumber\\ 
{\sigma_y} & = & .0201\ \mu{\rm{m}}
\label{tray17}
\end{eqnarray}

Just as in Eq.\ref{tray14}, the emittances in the vertical place, $\epsilon_y$, goes as Eq.\ref{tray18},

\begin{equation}
{\epsilon_y} = {{\sigma_y^2}\over{\beta^*_y}} = {{(.0201\ \mu{\rm{m}})^2}\over {(600\ \mu{\rm{m}})}} = 0.00067\ \rm{nm}.
\label{tray18}
\end{equation}

We calculated the $\sigma_x$, $\sigma_y$, $\epsilon_x$, and $\epsilon_y$ for the 200 GeV energy collision as well, utilizing the exact same procedure as above.  Those values can also be found in Table \ref{table:nonlin}.

One factor that must be taken into consideration is whether or not this emittance is even possible.  This is easily done by looking at our emittance projections compared to the current standards at the ILC.  The current values achievable at ILC for horizontal and vertical emittances are $\epsilon^n_x$ = 12 mm$\cdot$mrad and $\epsilon^n_y$ = .08 mm$\cdot$mrad \cite{Phinney}.
Converting these values using our proposed gamma factor for 250 GeV, $\gamma$,

\begin{equation}
\gamma = {{E_{collision}}\over{m_{e^-}}} = {{250\ \rm{GeV}}\over{511\ \rm{keV}}} = 489 000 
\label{tray19}
\end{equation}

Then converting mm$\cdot$mrad $\rightarrow$ nm, 

\begin{eqnarray}
{\epsilon_x} & = & {{\epsilon^n_x}\over{\gamma}} = {{12 \times 10^{-6}\ \rm{m}}\over{489000}} = 0.0245\ \rm{nm}\nonumber\\
{\epsilon_y} & = & {{\epsilon^n_y}\over{\gamma}} = {{0.08 \times 10^{-6}\ \rm{m}}\over{489000}} = 0.00016\ \rm{nm}
\label{tray20}
\end{eqnarray}

It is easy then to see by Eq.\ref{tray20} that the values we proposed are both larger than the values put forth by the current ILC parameter proposal [$\epsilon_x \rightarrow {{0.9\ \rm{nm}}\over{0.0245\ \rm{nm}}}\sim40,\ \epsilon_y \rightarrow {{0.00067\ \rm{nm}}\over{0.00017\ \rm{nm}}}\sim4$].  This is very inspiring because these projections are actually realistic when compared to the values from ILC that have already been accomplished with ultra low emittance damping ring\ \cite{Rubin}. 

When considering that the dipoles for this 233 km circumference ring have a magnetic 
field 4 times lower than used at the CERN LEP machine, to maintain good field quality, a soft magnetic material is needed\ \cite{Gourber1979,Gourber1981,Gourber1983,Laeger}. Grain oriented silicon steel is chosen for the dipoles because its coercivity is an order of magnitude lower than low carbon steel\,\cite{Shirkoohi}. Horizontal bands are added to the top and bottom of $C$ shaped laminations to permit high permeability path in the entire flux return circuit. Putting concrete in between laminations would provide space for the two bands.  An example of this design is currently being tested and calibrated at the University of Mississippi, Fig.\ref{magnet} \cite{Summers2012}.

The beam-beam bremsstrahlung cross section becomes a fair fraction of a barn, leading to $10^{35} \cdot 10^{-24} = 10^{11}$ interactions per second. With this high of an interaction frequency, the beam must be replaced every few minutes.  The horizontal tune shifts, $K_2$ strength of Crab Waist sextapoles, and maximum beam size in interaction point focusing optics are calculated as follows~\cite{Raimondi2007,Raimondi2001}:   

\begin{eqnarray}
{K_2} & = & {{1\over{2\theta_h}}{1\over{\beta_y \beta_y^*}}{\sqrt{{\beta_x^*}\over{\beta_x}}}}\nonumber\\
{K_2} & = & {{1\over{2(0.017)}}{1\over{(250000\ \rm{m})(0.0006\ m)}}{\sqrt{{0.02\ \rm{m}}\over{40000\ \rm{m}}}}}\nonumber\\
{K_2} & = & 0.0014
\label{tray100}
\end{eqnarray}

Which appears to be achievable, for a sextapole.  Now on to calculating the maximum beam sizes in the IP quadrupoles to see if they fit , Eq.\ref{tray101},

\begin{eqnarray}
{\sigma_x} & = & {\sqrt{\epsilon_x \beta_x}} = {\sqrt{0.9\times 10^{-9} \cdot 40000}} = 6\ \rm{mm}\nonumber\\
{\sigma_y} & = & {\sqrt{\epsilon_y \beta_y}} = {\sqrt{0.00067\times 10^{-9} \cdot 250000}} = 0.41\ \rm{mm}
\label{tray101}
\end{eqnarray}

Finally we can calculate the tune shifts in each plane\ \cite{Raimondi2006}, Eq.\ref{tray102},

\begin{eqnarray}
{\xi_x} & = & {{2r_eN\over{\pi \gamma}}{{\beta_x}\over{\sigma_z^2 \theta^2}}} = 0.027\nonumber\\
{\xi_y} & = & {{r_eN\over{\pi \gamma}}{{\beta_y}\over{\sigma_y \sigma_z \theta}}} = 0.23
\label{tray102}
\end{eqnarray}

Note that luminosity, RF power, and bunch length can be traded off. A
longer bunch length allows a lower injection energy by minimizing the
transverse mode coupling instability\,\cite{Sen}. The power for the
superconducting RF cavities might come from klystrons or from phase and
frequency locked magnetrons\,\cite{Neubauer,Dexter}.

\begin{figure}[ht]
  \centerline{\includegraphics[width=5in]{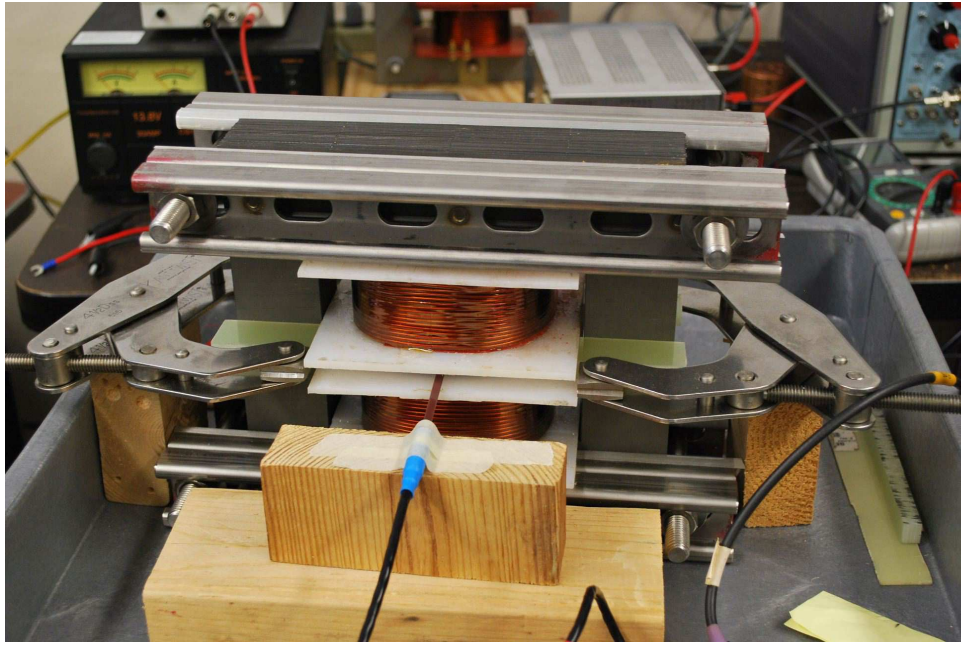}}
  \caption{1.8 Tesla, 400Hz Grain Oriented Silicon Steel Dipole Magnet at the University of Mississippi\ \cite{Summers2012}}
  \label{magnet}
 \end{figure}

\begin{table}[ht] 
\caption{$e^{+}e^{-}$ Collider Parameters}\smallskip
\centering   
\tabcolsep=1.2mm   
\begin{tabular}{l c c c c} 
\hline\hline\                      
Parameter & LEP & VLLC & $Crab Waist_{200}$ & $Crab Waist_{250}$\\ [0.5ex] 
\hline                    
Circumference (m) & 26 658.9 & 233 000.0 & 233 000.0 & 233 000.0  \\   
$\beta^*_{x}$, $\beta^*_{y}$ (cm) & 150, 5 & 100, 1 & 2, .06 & 2, .06  \\ 
Luminosity $(cm^{-2}$ $sec^{-1})$ & 9.73 $\times$ $10^{31}$ & 8.8 $\times$ $10^{33}$ & 1.5$\times$ $10^{35}$ & 9.7$\times$ $10^{34}$ \\ 
Energy (GeV) & 97.8 & 200.0 & 200.0 & 250.0\\ 
$\gamma$ & 191 000 & 391 000 & 391 000 & 489 000\\
Emittances $\epsilon_x$, $\epsilon_y$ (nm) & 21.1, 0.220 & 3.09, 0.031 & .9, .0017 & .9, .00067\\
rms beam size IP $\sigma^*_x$, $\sigma^*_y$ ($\mu$m) & 178.0, 3.30 & 55.63, 0.56 & 4.25, 0.0321 & 4.25, 0.0201\\ 
Bunch intensity/I (/mA) & ${4.01 \times 10^{11}}\over 0.720$ & ${4.85 \times 10^{11}}\over 0.1$ & ${4.85 \times 10^{11}}\over 0.1$ & ${4.85 \times 10^{11}}\over 0.04$\\ 
Number of bunches per beam & 4 & 114 & 114 & 46\\ 
Total beam current (mA) & 5.76 & 22.8 & 22.8 & 9.34\\ 
Beam-beam tune shift $\xi_{x}$, $\xi_{y}$ & 0.043, 0.079 & 0.18, 0.18 & 0.034, 0.18  & 0.027, 0.23\\ 
Dipole field (T) & 0.110 & 0.0208 & 0.0208 & 0.0260\\ 
E loss / $e^{\pm}$ / turn (GeV) & 2.67 & 4.42 & 4.42 & 10.8\\ 
Bunch length (mm) & 11.0 & 6.67 &  6.67 &  6.67\\ 
Revolution frequency (kHz) & 11.245 &  1.287 & 1.287 & 1.287\\ 
Synch rad pwr (b.b.) (MW) & 14.5 & 100.7 & 100.7 & 100.7\\  [1ex]       
\hline\hline     
\end{tabular} 
\label{table:nonlin}  
\end{table} 
\clearpage
\vfill

\mbox{}\vspace{.0in}\\
\section{Proton - Antiproton}

\subsection{Theory}
\label{sec:meaningfulname}
\indent

One of the most abundant sources of naturally occurring antiproton production occurs when cosmic ray protons interact with the interstellar medium.  This example will serve as a good model for the creation of antiprotons.  Essentially you have a very energetic proton traveling through space that collides with any arbitrary nucleus in the interstellar medium, and the following reaction occurs

\begin{equation}
p + A \rightarrow p + \bar{p} + p + A
\label{tray21}
\end{equation}

Where in Eq.\ref{tray21}, $p$ represents the proton, $A$ represents the interacting nucleus, and $\bar{p}$ represents the antiproton.  After the interaction all particles, including the antiproton, continue through space until colliding or reacting with other particles.

\subsection{The Tevatron}
\label{sec:meaningfulname}
\indent

Although CERN built the first machine with proton-antiproton collisions \cite{Rubbia}, the highest energy model came in the flagship design of the Tevatron at Fermilab.  Originally designed for fixed target collisions, the main collider ring at Fermilab up into the 70's could only accomplish orbiting energies of about 400 GeV. In 1977 however, Bob Wilson foresaw a way of constructing a new ring attached to the existing main ring that would double the then current maximum orbiting energy of an injected proton.  Essentially he wanted to construct a``Super Ring" that would implement the relatively new concept of magnets that accomplished superconductivity by supercooling.  In creating this Tevatron Super Ring, a proton could be ramped up to an orbiting energy of 1 TeV, then passed through the main collider ring to interact with another proton at 250 GeV \cite{Wilson}. 

After the completion of the Tevatron, initial applications came in ramping a proton to 800 GeV and smashing it into fixed targets.  Soon after CERN showed the possibility of sustaining and colliding antiprotons, however, the Tevatron had to switch gears. In 1985, Fermilab introduced the Antiproton Source into the collider design that could create antiprotons and send them into the Tevatron for collision with protons.  All collisions therefore would occur and be monitored in the Tevatron itself.  Vast improvements have been made in antiproton production and peak luminosities since that time.  Recently, peak luminosities were on the order of $3 \times 10^{32}\ \rm{ cm^{-2} sec^{-1}}$.  Just in the 5 year span after slip stacking was introduced in 2005~\ \cite{Shukla}, weekly antiproton production went up roughly 8 times ($\sim500 \times 10^{10}$ antiprotons/Week in 2005$\rightarrow$ $\sim4000 \times 10^{10}$ antiprotons/Week in 2009) \cite{Pasquinelli}.  The most current data from Fermilab (Fig.\ref{fermilab}) shows peak luminosity to be on the order of $4 \times 10^{32}\ \rm{ cm^{-2} sec^{-1}}$ \cite{Holmes} and this is the luminosity we will use to discuss design implementation into our proposed system in the next section.

\begin{figure}[ht]
  \centerline{\includegraphics[width=4.75in]{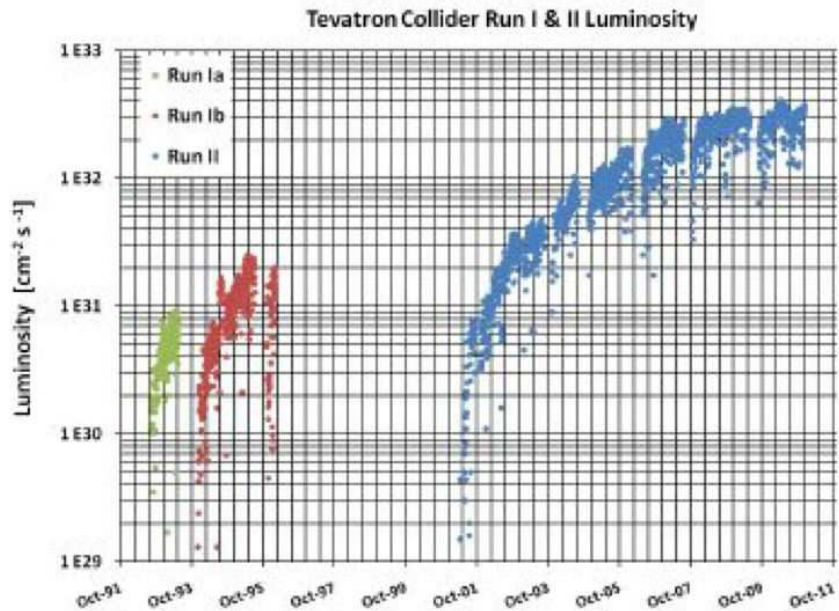}}
  \caption{Current Fermilab Luminosity Data\ \cite{Holmes}}
  \label{fermilab}
 \end{figure}
 \vfill\break
 
\subsection{Future Projections}
\label{sec:meaningfulname}
\indent

Although there are many implementations that can be made to improve upon the current Tevatron design, we will focus on the a few that could be applied to the implementation of a new collider, similar to the Tevatron, to be installed in the VLHC\ \cite{Ambrosio,Blaskiewicz}, tunnel that was the focus of Chapter 1.  One immediate concern that arises out of the VLHC design we've proposed, is the sheer circumference difference between the current Tevatron (6.28 km) and the VLHC we've proposed (233 km).  If the collider ring circumference was in fact 233 km, then collision frequency would go down roughly by a factor of 37 (${6.28 \over 233} \sim {1\over 37}$).  However, because we're talking about a final collision energy of 20 TeV versus the 1 TeV currently implemented in the Tevatron, you gain a factor of about a 20 times more energetic collision, thus your luminosity would scale roughly as 

\begin{eqnarray}
{Luminosity_{proposed} = {({Energy_{increased}} \times {Frequency_{decreased}}) \times Luminosity_{current}}}\nonumber\\
{Luminosity_{proposed} = {({20} \times {1\over 37}) \times({4 \times 10^{32}\ \rm{cm^{-2} sec^{-1}}})} = {2.16\times 10^{32}\ \rm{cm^{-2} sec^{-1}}}}
\label{tray22}
\end{eqnarray}

So it's easy to see that we've basically gone down by a factor of two when simply considering putting current designs, with a higher collision energy, into the 233 km collider ring.  This is a fairly high peak luminosity, but we can still do a bit better by increasing another factor, $\sigma$. Although proton-proton ($pp$) and proton-antiproton ($p\bar{p}$) collision are virtually indistinguishable for low mass objects as threshold mass objects are approached, as mass and energy increase, there is a verifiable difference in the behavior of the two interactions.  Simulating three scenarios including: $pp$ at LHC 7$\times$7, upgraded $pp$ at LHC 20$\times$20, and a proposed $p\bar{p}$ at VLHC 20$\times$20, you can see a vast difference in the behavior at higher mass collisions in Fig.\ref{ppbar} \cite{Duraisamy}.  See also Fig. Fig.\ref{mrenna} \cite{Mrenna}.

\begin{figure}[ht]
  \centerline{\includegraphics[width=4in]{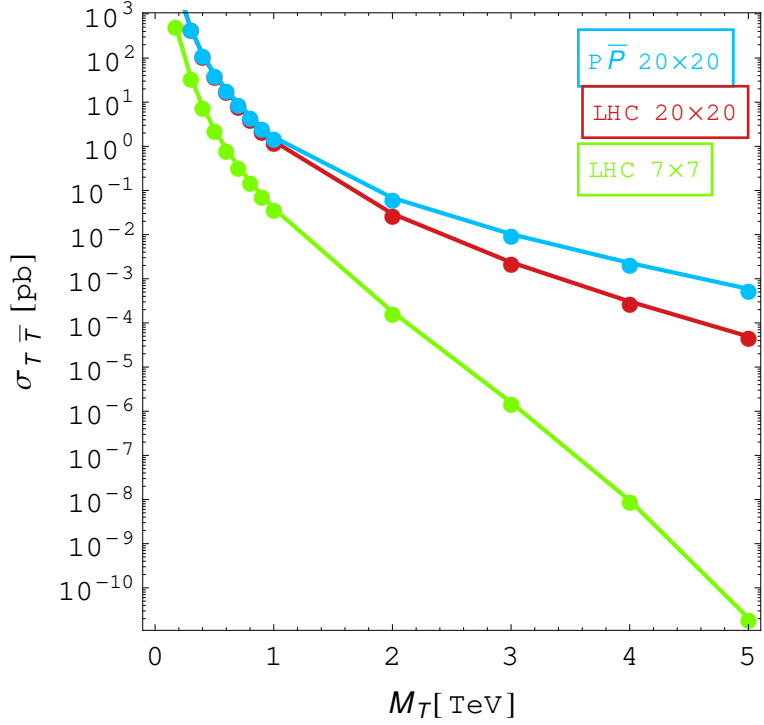}}
  \caption{Cross Section vs Collision Mass\ \cite{Duraisamy,Stelzer,Alwall2007,Alwall2011}}
  \label{ppbar}
 \end{figure}

Finally we see that when considering the current LHC parameters applied to $pp$ collisions compared to the proposed collisions of $p\bar{p}$, at higher mass collisions, $\sigma$ goes up by a factor of 10.   Most importantly, when considering the Event Rate, it must go up by a factor of ten as well ($\sigma \ \rightarrow 5 \times 10^{-4} \ \rightarrow 5 \times 10^{-3}$).  This works out great, as the Event Rate goes as,

\begin{eqnarray}
{Event \ Rate} & = & {(Luminosity_{proposed} \times \sigma)}\nonumber\\
{Event \ Rate} & = & {(Luminosity_{proposed} \times 10\sigma)}\nonumber\\
{{10(Event \ Rate)}} & = & {((2.16 \times 10^{32}\ \rm{cm^{-2} sec^{-1}}) \times 10\sigma)}
\label{tray23}
\end{eqnarray}

So then by Eq.\ref{tray23}, we clearly have a 10 times higher event rate when considering $p\bar{p}$ interactions at higher mass collisions compared to $pp$ for the same luminosities.  So we can see that interesting event rates are about the same, but the background collisions are lower.  Thus a lower cost detector with fewer channels is possible while keeping channel occupancy constant. 

So for high mass objects near theshhold, this collider, with the current
Fermilab $\bar{p}$ source, has double the event rate of the
Superconducting Super Collider (SSC) design. The current limitation on
the Fermilab $\bar{p}$ source is the accumulator ring which has a
$\bar{p}$ stacking rate of $26 \times 10^{10} \ \bar{p}$/hour.
The debuncher ring can supply $40 \times 10^{10} \ \bar{p}$/hour,
so a second $\bar{p}$ accumulator ring used in parallel might help.

\begin{figure}[ht]
  \centerline{\includegraphics[width=5in]{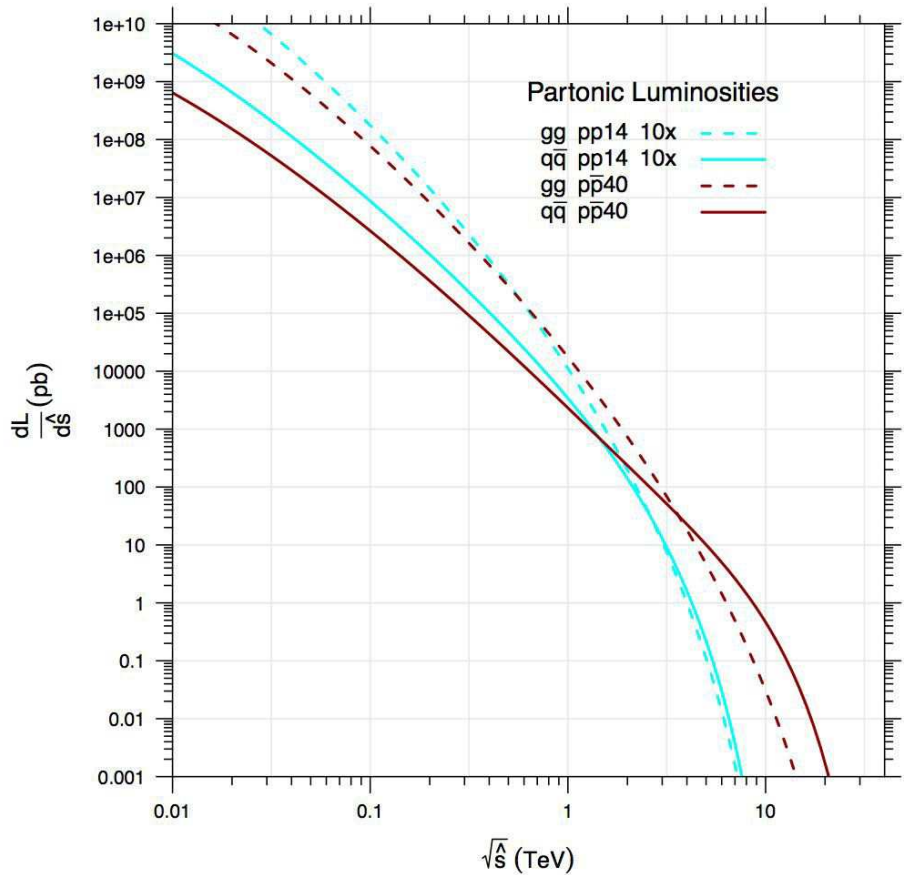}}
  \caption{Partonic Luminosity Data\ \cite{Mrenna}}
  \label{mrenna}
 \end{figure}
 \vfill\break
 
 \newpage
 \subsection{Collider Design}
\label{sec:meaningfulname}
\indent

To date there have been many particle collider built and functioning around the world.  Many more have been considered, including the Very Large Hadron Collider, or VLHC\ \cite{Ambrosio}.  Previous work at VLHC used roughly 100,000 amps.  This current came in the form of a magnet construction that entailed a half turn around two bores with a Niobium Titanium superconductor at T = 4K with liquid helium in an 8 cm diameter pipe.  

The design could be altered to bring the current consumption down by a factor of 4.  In the previous PIPETRON design (Fig.\ref{Pipetron_dipole}), a centered circular tube of current directs two paths for beam collision (Fig.\ref{Pipetron_dipole2}).  By switching to an H-frame dipole construction (Fig.\ref{frame8}), you would only have one beam path, thus cutting current by a factor of two, to 50,000 amps.  By implementing a full turn and one bore around the H-frame dipole magnet, only one offset current loop would be necessary, thus returning the return current back through the magnet.  This would buy you another factor of two, bringing current consumption to 25,000 amps.  With only one offset current loop in the H-frame dipole there could exist the possibility of a skew quadrupole field, but nevertheless, a lattice may still be possible \cite{Parker}.  

\begin{figure}[ht]
  \centerline{\includegraphics[width=3.75in]{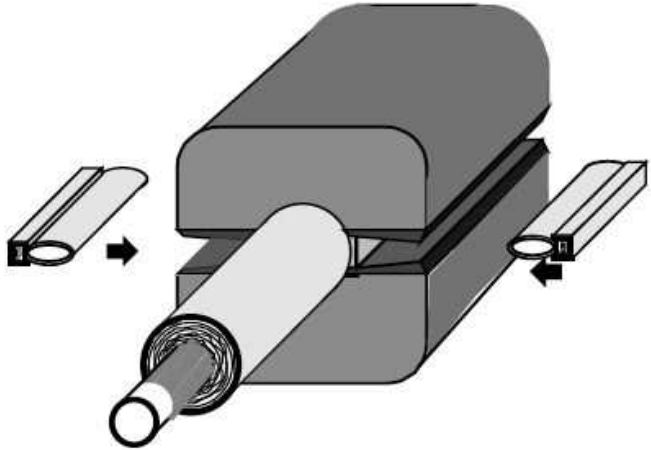}}
  \caption{Pipetron Schematic.\ \cite{Foster1997}}
  \label{Pipetron_dipole}
\end{figure} 

\begin{figure}[ht]
  \centerline{\includegraphics[width=2in]{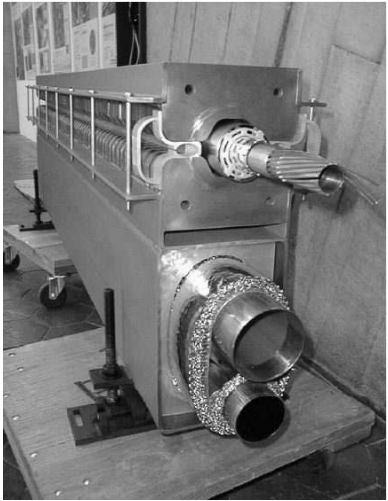}}
  \caption{Finished Pipetron Prototype.\ \cite{Foster1999,Piekarz}}
  \label{Pipetron_dipole2}
\end{figure} 
 
\begin{figure}[ht] 
  \centerline{\includegraphics[width=2.5in]{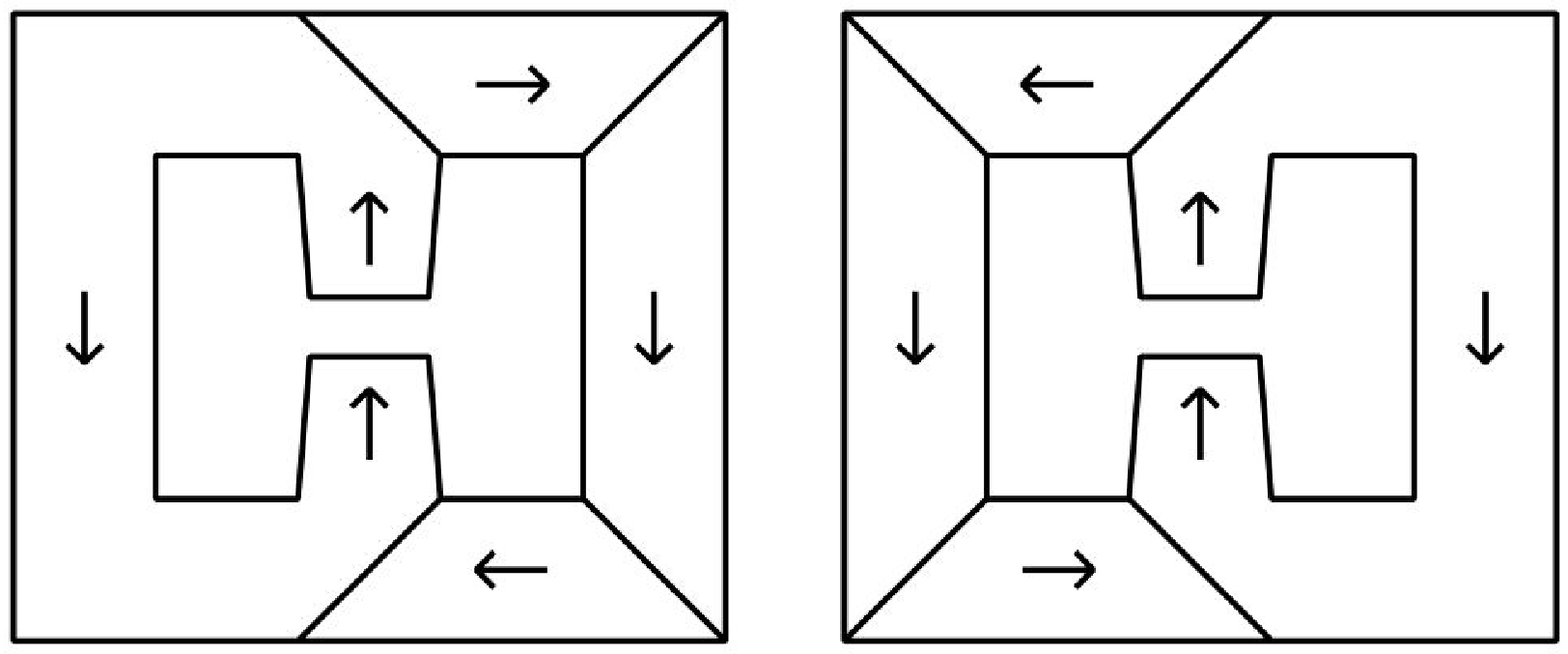}}
  \caption{H-frame Dipole Schematic\ \cite{Alexahin} (The mitered joints are needed for grain oriented silicon steel, but not with low carbon steel.)}
  \label{frame8}
\end{figure} 

\vfill\break

Another possible design change could come in the form of substituting SuperPower YBCO superconductors with T = 25K triple point liquid neon instead of liquid helium\ \cite{Lehner,Wu,Rossi}.  By implementing liquid neon and SuperPower YBCO, cooling costs could be minimized.  The availability of liquid neon is adequate.  When considering the heat of vaporization (HOV) of each element, neon has an HOV of approximately 1.71 kilo-Joule/mole, whereas helium has an HOV of approximately 0.083 kilo-Joule/mole.  This means that neon could absorb roughly 21 times as much heat per mole as helium in a quench, if the superconductor goes normal.  Also, considering that both gaseous helium and neon are lighter than air, neon might also be usable in a tunnel where oxygen deficiency is a concern.  So we should estimate the amount of neon needed to accomplish such a collider design. 

\begin{equation}
Volume_{collider} = 2 \times 233\ \rm{km} \times {\pi ( {.5\ \rm{cm}})^2} = 36.6 \ million \ cc
\label{tray24}
\end{equation}

Now if we assume that the liquid density of neon is 1.2 g/cc and the gaseous density of neon is 0.0009 g/cc, then the gaseous form is roughly 1,300 times less dense than the liquid form.  Then it is easy to see that collider ring would require 47.6 million liters of neon gas.  Linde currently sells neon gas for approximately 8 cents/liter, so the approximate cost of neon gas would come in at \$3.8 million.  This cost is pretty good, but the neon would necessitate recovery after quenches.  Linde currently produces about 300 million liters of gaseous neon per year.  Their delivery method requires full size tube trailers that hold 3.2 million liters each, so a project of this size would require a fleet of 15 Kenworth big trucks, each hauling a tube trailer to complete the delivery. 

Implementing the SuperPower YBCO Superconductors would be more challenging.  Currently $0.01\ \rm{mm} \times 12\ \rm{mm}$ 2G HTS ribbon costs \$70/meter and has the physical parameters to handle 1,500 amps at T = 25K.  To get the 25,000 amps needed for our proposed design, a minimum of 17 ribbons would be required.  Then the total length of YBCO Superconductor ribbon would amount to $ 17 \times 2 \times 233,000\ \rm{m} = 8,000,000 \ \rm{m}$.  At the proposed cost of \$70/meter, the total cost of the ribbon would be about \$555 million.  This cost seems a bit pricey and it may turn out that Niobium Titanium and liquid helium at T = 4K are cheaper, but the cryogenic load will be far greater at 4K than at 25K. 

\clearpage

\mbox{}\vspace{.0in}\\
\section{Muon - Muon}

\subsection{Theory}
\label{sec:meaningfulname}
\indent

Muons that are accelerated or stored in collider rings\ \cite{Neuffer1983,Neuffer1987,Summers1994,Fernow,Gallardo,Neuffer1997,Ankenbrandt,King1999,Alsharoa,Palmer2005,Summers2007,Palmer2007} will decay. The simplest form of muon decay comes in the form of Eq.\ref{tray25}, where a muon decays into an electron, an electron-antineutrino, and a muon-neutrino. The alternative decay is represented in the second half of Eq.\ref{tray25}, in which the antimuon decays into a positron, an electron-neutrino, and a muon-antineutrino.  These decays happen continuously and will eventually contribute an intense disk of neutrino radiation emanating tangentially out of the plane of a collider ring\ \cite{King1999B}.  

\begin{eqnarray}
{\mu^{-} \rightarrow  \nu_{\mu} + \bar{\nu_{e}} + e^{-} } \nonumber\\
{\mu^{+} \rightarrow \bar{\nu_{\mu}} + {\nu_{e}} + e^{+} }
\label{tray25}
\end{eqnarray}

Muons typically have a lifetime of 2.2 $\mu$s, however the muons in the scope of this project are boosted to energy levels on the order of 20 TeV, so their lifetimes are greatly increased.  The boost factor $\gamma$ is defined in Eq.\ref{tray26} and a sample calculation of the boosted mean lifetime is included as well.

\begin{eqnarray}
{\gamma} & = & {E_{\mu} \over m_{\mu} \times c^{2}} \nonumber\\
{\gamma} & = & {{{20\ \rm{TeV}} \over {106\ \rm{Mev}}} = 188,680} \nonumber\\
{\tau_{boosted}} & = & {{\tau_{\mu}} \times \gamma = 2.2\ {\mu}\rm{s} \times 188,680 = .415\ s}
\label{tray26}
\end{eqnarray}
 
A very direct concern comes in the size of the beam that ejects at the earth's surface.  Essentially this beam width can be approximated by examining the amount of energy each neutrino receives from the decay and what amount of that energy is contributed in the perpendicular ``kick" to the beam path direction.  If we approximated each particle in the resultant muon decay to receive a third of the 20 TeV muon directed energy, each component should travel with about 6.6 TeV.  In the case of the neutrino, the perpendicular kick should be about 20 Mev, so the appropriate opening angle is illustrated in Eq.\ref{tray27}.

\begin{eqnarray}
{\theta = {{Opposite} \over {Adjacent}} = {{20 \ \rm{Mev}} \over {6.6\ \rm{TeV}}} \simeq 3 \mu{\rm{radians}}}
\label{tray27}
\end{eqnarray}

\subsection{Past Research}
\label{sec:meaningfulname}
\indent

 Many reports have explored and approximated the harmful effects of a high energy muon collider and several more factors have to be considered and addressed.  One such concern is the amount of neutrino radiation dosage that exits at the earth's surface. One such approximation comes in the form of Eq.\ref{tray28}.\ \cite{King1999B}
 
 \begin{eqnarray}
{D_{exit}^{ave} [Sv]} = {2.9 \times 10^{-24}} \times {{{N_\mu} \times {({E_\mu} [\rm{TeV}])^3}} \over D[\rm{m}]}
 \label{tray28}
 \end{eqnarray}

 If we consider the collider ring to be 300 meters below ground (D = 300 m),  the energy of the muon to be approximately 20 TeV ( E = 20 TeV), and the constant to be relatively accurate, we only have one variable left to define.  The number of muons we need per year goes down greatly because their lifetimes have gone from roughly 2.2 $\mu$s to 0.415 seconds.  This means that the number of bunches and cycles goes down greatly because fewer bunches of muons are needed as the muons created simply live longer. If we consider 2 bunches of $2 \times 10^{12}$ muons per cycle, with each muon having a mean lifetime of 0.4 secs, then we arrive at about $3.1 \times 10^{20}$ muons per year.  Plugging all these parameters into Eq.\ref{tray28} will yield Eq.\ref{tray29}.

\begin{eqnarray}
{D_{exit}^{ave} [Sv]} = {2.9 \times 10^{-24}} \times {{{(3.1 \times 10^{20})} \times {({20}\ \rm{TeV)^3}}} \over 300\ \rm{m}} = {2400\ {\rm{[mRem/ year]}}} 
\label{tray29}
\end{eqnarray}

This yearly dosage is roughly 24 times more than the federal limit of approximately 100 mRem/year.   

\subsection{Future Projections}
\label{sec:meaningfulname}
\indent

Although there are several approaches to minimizing the yearly dosage of neutrino radiation, perhaps the most straightforward solution would come in simply spreading out the size of the radiated beam.  We must first consider a hypothetical circumstance in the present approximated beam with an opening angle described in the previous section of this paper.  If we approximate the beam surfacing at a distance of 250,000 feet away, the beam width that surfaced would be approximately 0.75 feet (9 inches).  This tightly concentrated beam would exceed allowable radiation limits, if considered to interact with a dense surface (concrete wall etc.), before showering a bystander, over a long period of time (all day, everyday, for years).  To spread out this concentrated beam may be a fairly simple fix. 

A roller coaster lattice is implemented in which the FODO (Focused-Off-Defocused-Off) quadropoles\ \cite{Courant} make the muon bunches oscillate up and down to offset the continuos neutrino decay only in a plane tangent to the track.  A variation of this design can be seen in the helical orbit focusing of protons and antiprotons implemented at the Fermilab Tevatron\ \cite{Goderre}.  Essentially, in the helical model you are continuously altering the path of the protons and antiprotons in their vertical and horizontal planes to have them spiral past one another without ever touching until they reach their specific collision site.  This helical model is a pretty straightforward addition to an existing collier ring and only requires adding electrostatic separators into the existing collider ring design. 

For muons, path oscillation would only occur in one plane, much like a roller coaster.  Simply introducing a roller coaster shaped muon stream instead of a planar circular track would not have any immediate advantage as the tangental neutrino decay would remain relatively the same.  The spreading of the otherwise tightly focused neutrino beam would come from introducing periodic bumps via short vertical dipoles into the track of the collider ring.  By adjusting these bumps at  random times, you could create a virtually random succession of phase shifted beams.  This phase shift would ultimately lead to a great spreading of neutrino beam interactions as it would spread out, at random, the vertical angle of the resultant neutrino decay at all points around the collider ring.  If the muons could rise 1 cm over 20 m, then an angle of 500 $\mu{\rm{radians}}$ could be achieved.  This would be roughly 160 times higher than the previous 3 $\mu{\rm{radians}}$ opening angle, more than curing the 24 times neutrino radiation dosage problem.  To put this in prespective, a dose of 15 [mRem/year] is equivalent to eating a banana at every meal.

\subsection{Collider Design}
\label{sec:meaningfulname}
\indent

When considering the construction of a 2 Tesla magnet for a 20 x 20 TeV muon collider in a 233 km long collider ring, previous work at VLHC used roughly 100,000 amps. If we consider the initial muon injection energy to be approximately 2 TeV and the energy gained per revolution of the entire collider ring to be approximately 250 GeV, then the approximate number of orbits to reach a overall energy of 20 TeV will be about 72.  If we take the muons to move near the speed of light, then we see that they make the 72 orbits in about .06 seconds.  This amounts to a 180 degree half cycle if an offset White circuit\ \cite{White} runs each magnet string, then the resultant frequency of the magnets becomes 8.3 Hz.  From previous research, many superconductors have been used on a small scale at up to 60 Hz\ \cite{Sumption,Auchmann,Fisher}.  Knowing that the eddy current losses go as the square of the frequency, the eddy current losses would be about 50 times less at 8.3 Hz as compared to 60 Hz.  

If we wanted to calculate the eddy current losses within the dipole magnets, such calculations are pretty straight forward\ \cite{Sasaki,Summers2007}.  Taking ultra low carbon steel 0.5 mm laminations to reach 2 Tesla, the dipoles ramp from 0.2 Tesla to 2 Tesla, for a swing of $\pm$ 0.9 Tesla.  Taking the volume of the steel to be $0.25\ \rm{m} \times 0.25\ \rm{m} \times 233,000\ \rm{m} = 15,000\ \rm{m^3}$ (roughly 2.5 times as much steel used in the LHC), we can calculate the power consumption of eddy current losses.
  
\begin{eqnarray}
{Power} & = & {{{Volume \times {( 2 \times \pi \times f \times B \times w)^2}} \over 24 \rho}} \nonumber\\
{Power} & = & {{{15,000 \times {( 2 \times \pi \times 8.3 \times .9 \times .0005)^2}} \over 24 \times (9.6 e^-9)}} \nonumber\\
{Power} & = & {36 \ \rm{Megawatts}}
\label{tray30}
\end{eqnarray}

But the duty cycle is roughly ${{0.12 \over 0.4} = 0.3}$ and not continuous, therefore the 36 Megawatts becomes about 11 Megawatts.  This consumption is well below the 200 Megawatts being discussed for other projects.  

Another power loss must be considered in the form of hysteresis losses.  If we consider the hysteresis losses in the dipoles which cycle around the BH loop every .4 secs, the energy goes as the area of the BH loop divided by $4\pi$ in CGS units. If we approximate the losses by a very straight forward method, the Steinmetz Law, we can estimate such losses\ \cite{Dawes,Lyman}. 

\begin{equation}
{{Energy/cycle} = {(.001 \ \rm{Oersted}) \times {( 9000^{1.6} \ Gauss)}} = 2100 \ \rm{ergs/cc}}
\label{tray31}
\end{equation}

then plugging Eq.\ref{tray31} in Eq.\ref{tray30} we get,

\begin{eqnarray}
{Power} & = & {{{(15,000 \ \rm{m^3}) \times (10^6 \ cc/m^3) \times (2100 \ ergs/cc) \times (10^{-7} \ ergs/joule) }} \over .4\ \rm{secs}} \nonumber\\
{Power} & = & {8 \ \rm{Megawatts}}
\label{tray32}
\end{eqnarray}

Again we're still well below a 200 Megawatt site power limit.  It also should be noted that substituting grain oriented silicon steel for very low carbon steel would lead to a large decrease in both eddy current and hysteresis losses.  But, in exchange, the maximum field would decrease from 2 Tesla to 1.8 Tesla, therefore leading to a 10\% lower beam energy. But losses in annealed 0.005\% ultra low carbon steel\,\cite{CMI,Bertinelli} appear to be low enough. The muon collider
magnets ramp quickly, so eddy current and hysteresis losses are a concern. The $p \bar{p}$ magnets ramp slowly.  One may want to ramp the $e^+ e^-$ magnets quickly to go from 20 to 44GeV in two orbits. This minimizes the time for the transverse mode coupling instability to cause trouble. The RF to do this is already in place and will be used at higher energies to make up for synchrotron radiation losses.

Now we can also see how the luminosity, $L$, scales with energy and collision frequency.  $L = f n N_1 N_2 / A$ where $f$ is the revolution frequency, $n$ is the number of bunches in one beam, $N_i$ is the number of particles in each bunch, and $A$ is the cross sectional area of the beam at the interaction point. Also it is important to note that $A = 4 \pi \sigma_x \sigma_y$, where $\sigma_x$ and $\sigma_y$ are transverse sizes. The transverse angular beam spread decreases linearly with energy as seen in $\sigma = \sqrt{\epsilon_N \beta^* / \beta \gamma}$ where $\epsilon$ is the normalized emittance in mm-mrad, $\beta^*$ is the collision region focal length, $\beta = v/c$, and $\gamma = 1/\sqrt{1 - \beta^2}$. So for constant focal length, $\beta^*$, the decrease in collision frequency in a larger ring is exactly canceled by the higher muon energy, which goes as $\gamma$, and the luminosity stays the same.  So at least then the luminosity doesn't go down. 

We must also consider the overall energy of the beam once it reaches the maximum muon energy of 20 TeV.  This calculation is also indicative of what the overall beam power would for the specific muon lifetime of $\sim$0.4 secs.  This specific lifetime essentially means that the muon injector must replenish the muon supply every 0.4 secs, giving it an operating frequency of approximately 2.5 Hz.
Then,
 
 \begin{eqnarray}
{Total \ Energy_{max}} & = & {2 \times (N_{Muons}) \times {( E_{max})} \times {\rm{Joules} \over eV}} \nonumber\\
{Total \ Energy_{max}} & = & {{2 \times (2 e^{12}) \times {( 20e^{12})} \times {1.6e^{-19}}} = 1.28 \times 10^7 \ \rm{Joules}}.
\label{tray33}
\end{eqnarray}

Then Power goes as,

\begin{eqnarray}
{Power} & = & {{E_{max}} \times f} \nonumber\\
{Power} & = & {{1.28 \times 10^7 \ \rm{Joules}} \times 2.5\ \rm{Hz}} \nonumber\\
{Power} & = & {32 \ \rm{Megawatts}}.
\label{tray34}
\end{eqnarray}

So it's easy to see that maximum power is still below the 200 Megawatt site limit, even when efficiency is added in.  The last calculation to make for the proposed collider design will be a model that replicates the muon survival rate when ramping up to the maximum energy of 20 TeV.  A few parameters must be defined, but essentially it acts as a product series of an exponential function. The muon survival rate can be defined as 

\begin{equation}
\hbox{SURVIVAL} = \prod_{N=1}^{Orbits} \exp\left[{{-2\pi{R}\,m} \over
{[{E_{injection}} + ({E_{orbit}}\,N)]\,c\tau}}\right] = Survival\%.
\label{tray35}
\end{equation}

So for the collider design proposed, we want to ramp up to 20 TeV, in 72 orbits.  We set the $E_{injection}$ to 2 TeV.  We also set the energy gained per orbit around the collider ring of the muon, $E_{orbit}$, to be .25 TeV.  Now considering we are talking about the proposed 233 km collider ring construction, the respective radius, $R$, would be 37,100 m.  Now we have all the parameters we need to evaluate the product summation of the series, but evaluating the series itself is not such a trivial calculation.  For this reason, I created a Mathematica program to evaluate this product.  The program is shown in Fig.\ref{trey_nb}.

Therefore evaluating $aval = 37100$ for the ring radius in meters, $bval = 2000$ for the $E_{injection}$ in GeV, and the $cval = 250$ for the $E_{orbit}$ in GeV, we calculated the survival rate to be approximately
 
\begin{equation}
\hbox{SURVIVAL} = \prod_{N=1}^{72} \exp\left[{{-2\pi{(37100)}\,m} \over
{[{2000} + ({250}\,N)]\,c\tau}}\right] = 71.58\%.
\label{tray36}
\end{equation}

\begin{figure}[ht]
  \centerline{\includegraphics[width=6in, height=4in]{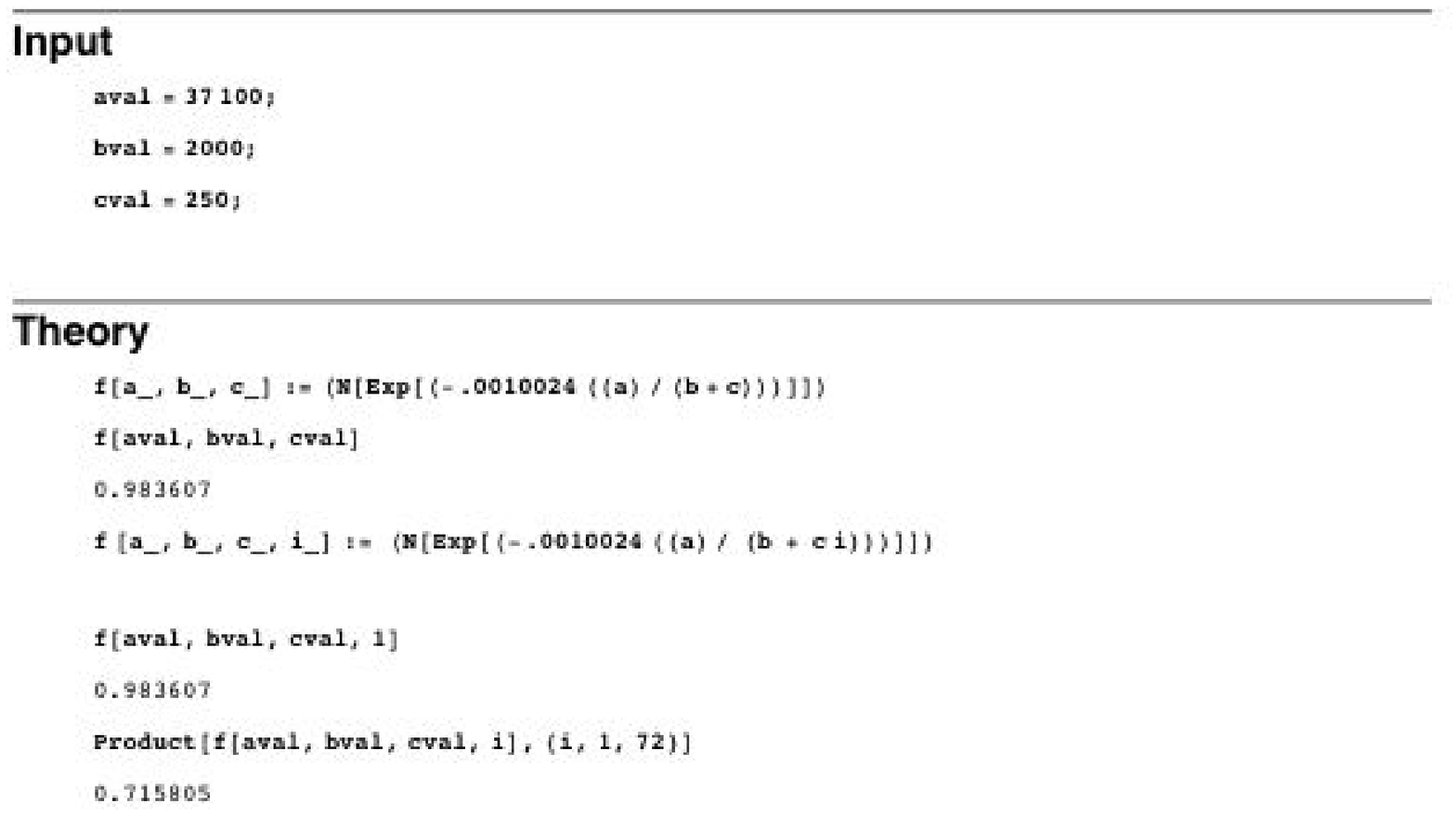}}
  \caption{Mathematica NoteBook Example.}
  \label{trey_nb}
\end{figure}

\clearpage

\mbox{}\vspace{.0in}\\
\section{Impacts on Society}

\subsection{Future Projections}
\label{sec:meaningfulname}
\indent

Robotic tunneling may be advanced, particularly in the areas of rock
removal\,\cite{Foster1998}, and shotcrete tunnel lining. Advancement of
the use of high temperature superconductors at 25 Kelvin may lead to
less expensive magnetic resonance imaging machines in medicine. It is
much easier to reach a temperature of 25 Kelvin than 4 Kelvin. The
multi-megawatt proton accelerators such as fixed field, alternating
gradient FFAG rings needed to produce muons may also be
able to fission thorium\,\cite{Bowman} to produce copious amounts of
electricity. Thorium reactor fission stops automatically in the
absence of the proton beam.

\clearpage

\mbox{}\vspace{.0in}\\
\section*{Conclusion}
\addcontentsline{toc}{section}{Conclusion}
\indent 

Ten years ago in 2001, the cost of a 233 km circumference tunnel in 
Northern Illinois was estimated to cost \$2.8 billion including 30\% 
contingency. Since then inflation has increased prices but more 
automation has been added to tunneling in pulling tunnel boring machines 
forward and in placing rock stabilization bolts.  This leads me to believe that construction costs would be directly on point with figures proposed over a decade ago, and perhaps even a little lower.

A 500 GeV center of mass $e^+ e^-$ collider with a luminosity of $10^{35}$
cm$^{-2}$ s$^{-1}$ appears to be possible. A crab waist crossing as developed
for the next generation B factories is employed and shows great promise for high luminosity by exploiting low emittance. Low emittance bunches from precision damping rings for the proposed
International Linear Collider (ILC) are used as well the short vertical focal
length ILC collision region optics. Forty times less installed RF acceleration 
is required in the ring as compared to a linear collider. Low coercivity grain oriented silicon steel is used for dipole magnets.

A 40 TeV $p \bar{p}$ fits in the tunnel with 2 Tesla H-frame iron 
dipoles. For many high-mass objects, the $p \bar{p}$ cross section is an 
order of magnitude larger than the $p p$ cross section. Single bore iron 
dipoles permit high temperature superconductor ribbons running at liquid 
neon temperatures. Advances have been made in $\bar{p}$ production at 
Fermilab. A detector with more physics and fewer background interactions 
per crossing requires fewer channels and therefore as a consequence costs less.

A $\mu^+ \mu^-$ collider approaching a center of mass energy of roughly 40 TeV fits in a 
233 km circumference tunnel. The same dipoles as used for $p \bar{p}$ 
can be employed. Eddy current and hysteresis losses are shown to be reasonable. 
The ramping rate is low enough so that superconductor can be employed. 
With 250 GeV of RF the muons are accelerated in 72 orbits. Muon survival 
$\sim$ 72\%. A phase shifting, FODO, roller coaster lattice similar to the 
Tevatron helical lattice is used to mitigate the neutrino radiation problem down 
to a level equivalent to eating one banana per meal. The vertical spread 
of the neutrinos is simply increased. Half the RF of the ILC is used to 
produce a lepton collider with 80 times the center of mass energy of the 
ILC, if muons can be cooled\,\cite{Palmer2007} at a reasonable cost\,\cite{Hart, Bao}.  Beam power for the muon and electron machines\,\cite{Phinney} is similar, 32 Megawatts vs. 21.6 Megawatts, respectively.

\clearpage

\section*{}
\addcontentsline{toc}{section}{REFERENCES}
\mbox{}\vspace{.15in}\\
\centerline{\bf REFERENCES}

\clearpage

\begin{singlespace}

\end{singlespace}

\clearpage

\section*{}
\addcontentsline{toc}{section}{VITA}
\mbox{}\vspace{.15in}\\
\centerline{\large \bf VITA}\\

\begin{itemize} 

\item George Thomas Lyons III is born. (1987)

\item Graduates from Mississippi School for Mathematics. (2005)

\item Enrolls at University of Mississippi. (2005)

\item Works with Dr. Joel Mobley at National Center for Physical Acoustics on ``Ultrasonic Scattering in Dispersive Micelles". (2007)

\item Works with Dr. Glenn Holt at Boston University on``Acoustic Levitation of Spherical Thrombus Sampling". (2008)

\item Inducted into Sigma Pi Sigma, serves two years as President and Congressional Representative for University of Mississippi. (2008)

\item Graduates from University of Mississippi with B.S. in Physics and B.A. in Mathematics with cumulative GPA of 3.3. (2009)

\item Graduates from University of Mississippi with a Master of Science in Physics and a cumulative GPA of 3.5. (2011)

\end{itemize}


\begin{thebibliography}{99}

\bibitem{Anjos}
J. C. Anjos {\it et al.,} ``Measurement of the Form Factors in the Decay $D^+ \to \bar{K}^{*0} e^+ \nu_e$," Phys. Rev. Lett. {\bf 65} (1990) 2630.
          
\bibitem{Aitala}
E. M. Aitala {\it et el.,} ``Correlations Between $D$ and $\bar{D}$ Mesons Produced in 500 GeV/c  $\pi^-$-Nucleon Interactions,"          
Eur. Phys. J. direct {\bf C1} (1999) 4.

\bibitem{Abazov2009}
V. M. Abazov {\it et al.,} ``Observation of Single Top Quark Production," Phys. Rev. Lett. {\bf 103}  (2009) 092001.      

\bibitem{Abazov2009B}
V. M. Abazov {\it et al.,} ``Measurement of the W Boson Mass," Phys. Rev. Lett. {\bf 103}  (2009) 141801.      

\bibitem{Barger}
V.\,Barger,
``Overview of Physics at a Muon Collider,"
AIP Conf.\,Proc. {\bf 441} (1997) 3.

\bibitem{Eichten}
E. Eichten, ``Towards a Muon Collider,"
ICFA Beam Dyn. Newslett. {\bf 55} (2011) 12.

\bibitem{CNA}
CNA Consulting Engineers (Hatch-Mott-MacDonald),
``Estimate of Heavy Civil Underground Construction Costs for a Very Large Hadron Collider in Northern Illinois,"
VLHC-2001-CNA-REPORT, http://vlhc.org/cna/cna\_report.pdf

\bibitem{Lach}
J.~Lach, R.~Bauer, P.~Conroy, C.~Laughton, and E.~Malamud, ``Cost Model for a 3 TeV VLHC Booster Tunnel,"
Fermilab\,-TM-2048, 1998.

 \bibitem{Marks}  
Theodore Baumeister and Lionel S. Marks,   
``Standard Handbook for Mechanical Engineers, 7th Edition," 1967,
ISBN-9780070041226, pp. 5-65.

 \bibitem{Schopper}
Herwig Schopper,
``Status of LEP and Its Experimental Programme,"
IEEE Trans. Nucl. Sci. {\bf 32} (1985) 1561.

\bibitem{Sen}
T. Sen and J. Norem, ``A Very Large Lepton Collider in the VLHC Tunnel,"
Phys. Rev. ST Accel. Beams {\bf 5} (2002) 031001.

\bibitem{Raimondi2007}
Pantaleo Raimondi, Dmitry N. Shatilov, and Mikhail Zobov,
``Beam-Beam Issues for Colliding Schemes with Large Piwinski Angle and Crabbed Waist,"
physics/0702033.

\bibitem{Zobov}
M. Zobov {\it et al.,} ``Test of ``Crab-Waist" Collisions at DA$\Phi$NE $\Phi$ Factory,"
Phys. Rev. Lett. {\bf 104}  (2010) 174801.

\bibitem{Shatilov}
Dmitry Shatilov, Eugene Levichev, Evgeny Simonov, and Mikhail Zobov,
``Application of Frequency Map Analysis to Beam-Beam Effects Study in Crab Waist Collision Scheme,"
Phys. Rev. ST Accel. Beams {\bf 14} (2011) 014001.

\bibitem{Raimondi2006}
P. Raimondi, ``Status on Super-B Effort,"
3rd Workshop on Super Flavor Factory based on Linear Collider Technology
(Super B III), Menlo Park, CA, 14-16 Jun 2006, pp.~104.

\bibitem{Phinney}
Nan Phinney, 
``ILC Reference Design Report: Accelerator Executive Summary,"
SLAC-PUB-13044, 2007.

\bibitem{Rubin}
David Rubin, ``The Challenges of Ultra-Low Emittance Damping Rings,"  
IPAC-2011-TUYB02, http://accelconf.web.cern.ch/AccelConf/IPAC2011/papers/tuyb02.pdf

\bibitem{Gourber1979}
J. P. Gourber and  L. Resegotti,
``Implications of the Low Field Levels in the LEP Magnets," 
IEEE Trans. Nucl. Sci. {\bf 26} (1979) 3185.

\bibitem{Gourber1981}
J. P. Gourber and  C. Wyss,
``Prototypes of LEP Bending Magnets with Steel-Concrete Cores,"
IEEE Trans. Nucl. Sci. {\bf 28} (1981) 2867.

\bibitem{Gourber1983}
J. P. Gourber, J. Billan, H. Laeger, A. Perrot, and L. Resegotti,
``On the Way to the Series Production of Steel Concrete Cores for the LEP Dipole Magnets,"
IEEE Trans. Nucl. Sci. {\bf 30} (1983) 3614.

\bibitem{Laeger}
H. Laeger, F. Beco, S. Comel, J. Hostettler, R. Luthi, and L. Vuffray,
``Production of the Soft Magnetic Steel Laminations for the LEP Dipole Magnets,"
IEEE Trans. Magnetics {\bf 24}  (1988) 835.

\bibitem{Shirkoohi}
G. H. Shirkoohi and M. A. M. Arikat,
``Anisotropic Properties of High Permeability Grain-Oriented 3.25\% Si-Fe Electrical Steel,"
IEEE Trans. Magnetics {\bf 30} (1994) 928.

\bibitem{Summers2012}
D. J. Summers {\it et al.,} ``Test of a 1.8 Tesla, 400 Hz Dipole for a Muon Synchrotron,"
IPAC-2012, New Orleans.

\bibitem{Raimondi2001}
Pantaleo Raimondi and  Andrei Seryi,
``A Novel Final Focus Design for Future Linear Colliders,"
Phys. Rev. Lett. {\bf 86} (2001) 3779.

\bibitem{Neubauer}
M. Neubauer {\it et al.,} 
``Phase and Frequency Locked Magnetrons for SRF Sources,"
IPAC-2011-MOPC140,   \newline
 http://accelconf.web.cern.ch/AccelConf/IPAC2011/papers/mopc140.pdf   
   
 \bibitem{Dexter}
 A. C. Dexter {\it et al.}
 ``First Demonstration and Performance of an Injection Locked Continuous Wave Magnetron to Phase Control a Superconducting Cavity,"   
 Phys. Rev. ST Accel.\,Beams {\bf 14} (2011) 032001.   

\bibitem{Rubbia}
C. Rubbia, P. McIntyre, and D. Cline,
``Producing Massive Neutral Intermediate Vector Bosons with Existing Accelerators,"
International Neutrino Conference, Aachen, Germany, 8-12 Jun 1976, pp. 683.

 \bibitem{Wilson}
Robert R. Wilson, ``The Tevatron," Phys. Today {\bf 30N10} (1977) 23.

\bibitem{Shukla}
S. Shukla, J. Marriner and  J. Griffin,
``Slip Stacking in the Fermilab Main Injector,"
SNOWMASS-1996-ACC015.

 \bibitem{Pasquinelli}
 Ralph J. Pasquinelli {\it et al.,} ``Progress in Antiproton Production at the Fermilab Tevatron Collider,"
 PAC09\,-Vancouver,  Fermilab-Conf-09-126-AD. 

\bibitem{Holmes}
Stephen Holmes, Ronald S. Moore, and Vladimir Shiltsev,
``Overview of the Tevatron Collider Complex: Goals, Operations and Performance,"
JINST {\bf 6} (2011) T08001.

\bibitem{Ambrosio}
Giorgio Ambrosio {\it et al.,} ``Design Study for a Staged Very Large Hadron Collider," Fermilab\,-TM-2149, 2001.

\bibitem{Blaskiewicz}
Mike Blaskiewicz {\it et al.,} ``Very Large Hadron Collider Instability Workshop: Summary Report," SLAC-PUB-8800, 2001.

\bibitem{Duraisamy}
Murugeswaran Duraisamy, personal communication, 2011.

\bibitem{Stelzer}
T.  Stelzer and W.  F. Long,
``Automatic Generation of Tree Level Helicity Amplitudes,"
Comput. Phys. Commun. {\bf 81} (1994) 357.

\bibitem{Alwall2007}
Johan Alwall {\it et al.,}
``MadGraph/MadEvent v4: The New Web Generation,"
JHEP {\bf 0709} (2007) 028.

\bibitem{Alwall2011}
Johan Alwall {\it et al.,}
``MadGraph 5:  Going Beyond,"
JHEP {\bf 1106} (2011) 128.

\bibitem{Mrenna}
Steve Mrenna, personal communication, 2011.

\bibitem{Parker}
B. Parker, ``Skew Quadrupole Focusing Lattices and Applications,"  PAC-2001-MOPA010.

\bibitem{Foster1997}
G. William Foster, P. O. Mazur, T. J. Peterson, and C. D. Sylvester,
``Design and Operation of an Experimental Double C Transmission Line Magnet for the Pipetron,"
PAC-1997, pp. 3392, \newline
http://accelconf.web.cern.ch/AccelConf/pac97/papers/pdf/6P002.PDF

\bibitem{Foster1999}
G. W. Foster {\it et al.,} 
``Conductor Design for the VLHC Transmission Line Magnet,"
PAC-1999-New York, pp. 182.

\bibitem{Piekarz}
Henryk Piekarz {\it et al.,}
``A Test of a 2 Tesla Superconducting Transmission Line Magnet System,"
IEEE Trans. Appl. Supercond. {\bf 16} (2006) 342.

\bibitem{Alexahin}
Y. Alexahin and D. J. Summers.
``Rapid-Cycling Synchrotron with Variable Momentum Compaction,"
 IPAC-2010-MOPE085, \newline
http://accelconf.web.cern.ch/AccelConf/IPAC10/papers/mope085.pdf

\bibitem{Lehner}
Traute F. Lehner (SuperPower), ``Development of 2G HTS Wire for Demanding Electric Power Applications,"
New Materials for Energy, Santiago de Compostela, Spain, 20-21 Jun 2011,
\newline http://www.superpower-inc.com/system/files/2011\_0620+ENERMAT+Spain\_TL+Web.pdf
  
\bibitem{Wu}
 M. K. Wu {\it et al.,} ``Superconductivity at 93 K in a New Mixed-Phase Y-Ba-Cu-O Compound System at Ambient Pressure," 
 Phys. Rev. Lett. {\bf 58} (1987) 908.   
    
 \bibitem{Rossi}
Lucio Rossi,
``Superconductivity: Its Role, Its Success, and Its Setbacks in the Large Hadron Collider of CERN,"
Supercond. Sci.Technol. {\bf 23} (2010) 034001.

\bibitem{Neuffer1983}
D. Neuffer, ``Principles \& Applications of Muon Cooling," Part.\,Accel. {\bf 14} (1983) 75.

\bibitem{Neuffer1987} 
D. Neuffer, ``Multi-TeV Muon Colliders," AIP Conf.~Proc. {\bf 156} (1987) 201.

\bibitem{Summers1994}
D. J. Summers, ``The Top Quark, the Higgs Boson, and Supersymmetry at $\mu^+ \mu^-$ Colliders,"
Bull. Am. Phys. Soc. {\bf 39} (1994) 1818.

\bibitem{Fernow}
R. C. Fernow and  J. C. Gallardo,
``Muon Transverse Ionization Cooling: Stochastic Approach,"
Phys. Rev. {\bf E52} (1995) 1039.

\bibitem{Gallardo}
J. C. Gallardo {\it et el.,} 
``Muon Muon Collider: Feasibility Study," Snowmass\,-1996, BNL-52503.

\bibitem{Neuffer1997}
D. Neuffer,
``Analyses and Simulations of Longitudinal Motion in $\mu$-Recirculating Linacs,"
Nucl. Instrum. Meth. {\bf A384} (1997) 263.

\bibitem{Ankenbrandt}
C. M. Ankenbrandt {\it et al.,} 
``Status of Muon Collider Research and Development and Future Plans,"
Phys.~Rev.~ST Accel.~Beams {\bf 2} (1999) 081001.

\bibitem{King1999}
Bruce J. King, ``Discussion on Muon Collider Parameters at Center of Mass Energies from 0.1 TeV to 100 TeV," physics/9908016.

\bibitem{Alsharoa}
M.  Alsharo'a {\it et al.,} 
``Recent Progress in Neutrino Factory and Muon Collider Research within the Muon Collaboration,"
Phys.~Rev.~ST Accel.~Beams {\bf 6} (2003) 081001.

\bibitem{Palmer2005}
R.\,B.\,Palmer {\it et al.,}  ``Ionization Cooling Ring for Muons,"
Phys.\,Rev.\,ST Accel.~Beams {\bf 8} (2005) 061003.

\bibitem{Summers2007}
D. J. Summers {\it et al.,}  ``Muon Acceleration to 750 GeV in the Tevatron Tunnel for a 1.5 TeV $\mu^+ \mu^-$ Collider,"
PAC07-Albuquerque, arXiv:0707.0302.

\bibitem{Palmer2007}
R. B. Palmer {\it et al.,}  ``A Complete Scheme of Ionization Cooling for a Muon Collider,"
PAC07-Albuquerque, arXiv:0711.4275.

\bibitem{King1999B}
Bruce J. King, ``Potential Hazards from Neutrino Radiation at Muon Colliders,"  physics/9908017.

\bibitem{Courant}
E. D. Courant and  H. S. Snyder,
``Theory of the Alternating Gradient Synchrotron," 
 Annals Phys. {\bf 3} (1958) 1.

\bibitem{Goderre}
 G. P. Goderre and  E. Malamud
``Helical Orbit Studies in the Tevatron," 
PAC\,-1989,
http://accelconf.web.cern.ch/AccelConf/p89/PDF/PAC1989\_1818.pdf

\bibitem{White}
M. G. White, F. C. Shoemaker, and G. K. O'Neill,
``A 3-BeV High Intensity Proton Synchrotron,"
CERN Symposium on High-Energy Accelerators and Pion Phyiscs, Geneva, Switzerland, 11-23 Jun 1956, pp. 525.

\bibitem{Sumption}
M. D. Sumption, E. W. Collings,  and P. N. Barnes, 
``AC Loss in Striped (Filamentary) YBCO Coated Conductors Leading to Designs for High Frequencies and Field-Sweep Amplitudes," 
 Supercond. Sci.Technol. {\bf 18} (2005) 1502.

\bibitem{Auchmann}
B.\,Auchmann, R.\,deMaria, and S.\,Russenschuck,
``Calculation of Field Quality in Fast Ramping Superconducting Magnets,"
IEEE Trans.\,Appl.\,Supercond. {\bf 18} (2008) 1569.

\bibitem{Fisher}
E. Fischer {\it et al.,}
``Design and Test Status of the Fast Ramped Superconducting SIS100 Dipole Magnet for FAIR,"
IEEE Trans.\,Appl.\,Supercond. {\bf 21} (2011) 1844.

\bibitem{Sasaki}
H. Sasaki, 
``Magnets for Fast Cycling Synchrotrons,"
KEK-PREPRINT-91-216.

\bibitem{Dawes}
Chester Laurens Dawes,
``A Course in Electrical Engineering, Volume 1: Direct Currents,"   
McGraw Hill, 1920,  ASIN-B000LCA2D2, pp. 182-183.

\bibitem{Lyman}
Taylor Lyman,
``Metals Handbook, Properties and Selection, Volume 1, 8th Edition,"
ISBN-0012983012, 1961, pp. 789.

\bibitem{CMI}
http://www.cmispecialty.com/31558\_CMI-B\_Data\_Sheet.pdf

 \bibitem{Bertinelli}
F. Bertinelli {\it et al.,}
``Production of Low-Carbon Magnetic Steel for the LHC Superconducting Dipole and Quadrupole Magnets,"
IEEE Trans. Appl. Supercond.  {\bf 16} (2006) 1777.

\bibitem{Foster1998}   
G. William Foster.
``An Unmanned, Battery-Operated TBM Tunneling Scenario for the 34\,km VLHC Injector Tunnel at Fermilab,"
Oct 1998,  \newline
http://www.w2agz.com/Documents/battery\_tunneling.pdf   
   
\bibitem{Bowman}
Charles D. Bowman and Rolland P. Johnson,
``Accelerators for Subcritical Molten Salt Reactors,"
IPAC-2011-THOAB01, \newline
http://accelconf.web.cern.ch/AccelConf/IPAC2011/papers/thoab01.pdf

\bibitem{Hart}
T. L. Hart, D. J. Summers, and K. Paul, ``Magnetic Field Expansion Out of a Plane: Application to Cyclotron Development,"
PAC-2011-WEP190.

\bibitem{Bao}
Y. Bao, A. Caldwell, and D. Greenwald,
``A Sphere Cooler Scheme for Muon Cooling,"
PAC-2011-MOP017, \newline
http://accelconf.web.cern.ch/AccelConf/PAC2011/papers/mop017.pdf


\end{thebibliography}
\end{document}